\documentclass[11pt,a4paper]{article}
\usepackage[T1]{fontenc}
\usepackage{lmodern,amsmath,amssymb,mathtools,microtype}
\usepackage[margin=24mm]{geometry}
\usepackage{graphicx,booktabs,placeins}
\usepackage[hidelinks]{hyperref}
\hypersetup{pdftitle={Polyakov loop model with exact static quark determinant in the 't Hooft--Veneziano limit: the SU(N) case},pdfauthor={S. Voloshyn}}
\newcommand{\Tr}{\operatorname{Tr}}
\newcommand{\Res}{\operatorname{Res}}

\newcommand{\dd}{\mathrm d}

\newcommand{\Ree}{\operatorname{Re}}
\newcommand{\Imm}{\operatorname{Im}}

\newcommand{\gp}{g_+}
\newcommand{\gm}{g_-}
\allowdisplaybreaks[2]
\title{Polyakov loop model with exact static quark determinant
in the 't Hooft--Veneziano limit: the $SU(N)$ case}
\author{S.~Voloshyn\\[3pt]
\small Bogolyubov Institute for Theoretical Physics, National Academy of Sciences of Ukraine,\\[-1pt]
\small 03143 Kyiv, Ukraine}
\date{}
\begin{document}
\maketitle
\begin{abstract}
We study the $SU(N)$ Polyakov loop model with the exact static quark
determinant in the 't Hooft--Veneziano limit, with $N_f/N=\kappa$
fixed. The mean-field problem reduces to a unitary matrix integral
with two independent sources and a determinant sector. We solve
the local saddle by the resolvent method and use its moments to
determine the phase diagram and thermodynamic observables. At real
chemical potential the model has a zero-density phase, an intermediate
phase with partial baryon filling, and a saturation phase. We obtain
explicit equations for both critical boundaries and simple formulas
in the massless, heavy-dense and dilute-flavor limits. In particular,
the two boundaries merge at $|\mu|=m$ as $\kappa\to0$, producing a
first-order threshold in the free energy normalized by $NN_f$.
We calculate the Polyakov loops, baryon density, chiral condensate
and two spatial screening masses. At imaginary chemical potential
we resolve the Roberge--Weiss period by keeping $N\,\mathrm{Im}\,\mu$
fixed, and compare the large-$N$ theory with finite-$N$ mean field
and spatial $SU(3)$ simulations of the same static action.
The algebraic structure of the local solution and its projective
transformations are given in the appendices.
\end{abstract}

\section*{Introduction}
\addcontentsline{toc}{section}{Introduction}

Effective Polyakov loop models provide a useful description of
thermal lattice gauge theory at nonzero chemical potential. The
spatial gauge links and the quark fields are integrated out, leaving
an interaction between the Polyakov loops and a local contribution
from the quark determinant. The resulting models are simpler than
the underlying gauge theory, but retain its center structure and
the competition between the quark mass and the chemical potential.
They also admit dual formulations which can be simulated without a
sign problem in regimes where the original Boltzmann weight is
complex \cite{BCVdual,DualPhase,DualMass}.

An important distinction concerns the treatment of the determinant.
Keeping only its term linear in the Polyakov loop is appropriate
at small quark fugacity. This gives a Gross--Witten--Wadia
(GWW) matrix integral \cite{GrossWitten,Wadia,BCVlarge}.
The exact static determinant contains higher winding contributions
and a finite number of quark states. In particular, it permits a
saturation phase which cannot be described by the linear
approximation. The approximation made here is to neglect spatial
quark propagation, while retaining the temporal determinant exactly.

At zero spatial coupling the model reduces to one-dimensional QCD.
Its large-$N$ phase diagram contains three regions
\cite{QCDone}. At sufficiently large mass and small chemical
potential the baryon density is zero. It becomes nonzero at a
lower critical boundary and grows through an intermediate phase,
before reaching its maximal value at an upper boundary. Once
saturation is reached, the free energy is linear in the chemical
potential and its mass derivative vanishes. A central question
of the present work is how this structure changes when the
Polyakov loops interact.

We consider the joint limit $N,N_f\to\infty$ with
$\kappa=N_f/N$ fixed. In the homogeneous large-$N$ mean-field
description, the interaction is replaced by two self-consistent
sources. The distinction between $U(N)$ and $SU(N)$ remains
essential in this limit. For $SU(N)$ the determinant constraint
introduces integer sectors, and their relative weights depend
on the chemical potential. The sector selected at equilibrium
determines the baryon filling. The corresponding $U(N)$ problem
and an earlier analysis of the $SU(N)$ model were considered in
Refs.~\cite{VolU,VolSU}; related large-$N$ phase and screening
calculations are given in Refs.~\cite{BCVphase,BCVscreening}.

Our calculation separates the local matrix integral from the
spatial self-consistency equations. We first solve the integral
with independent sources and a fixed determinant sector.
Its two first moments then determine the mean field, while
the extremum over the sector gives the density. The resolvent
method gives a finite set of algebraic endpoint equations.
The thermodynamic thresholds are considerably simpler than the
general local solution: the lower boundary reduces to a quadratic
equation after a reciprocal change of coordinate, and the upper
boundary is cubic. These formulas allow us to follow the two
boundaries without fitting the free energy or extrapolating a
small-source series.

Several limits make the physical content particularly transparent.
The massless theory has no positive lower threshold, and its
saturation boundary is explicit. In the heavy-dense limit one
retains the quark fugacity while the antiquark fugacity disappears.
In the opposite flavor limits the free energy approaches the
threshold function $\max(m,|\mu|)$ as $\kappa\to0$, or the
free static-fermion expression
$\log(2\cosh m+2\cosh\mu)$ as $\kappa\to\infty$.
Throughout this paper the thermodynamic free energy is defined
with the sign of $\log Z$ and normalized per site and per $NN_f$,
as in Refs.~\cite{QCDone,VolU,VolSU}.

The local solution also determines the spatial correlation functions.
Their leading connected contribution follows by linearizing the
Schwinger--Dyson equation about the mean field. In doing this the
local determinant sector must be allowed to relax when a source
is varied. This gives two screening channels. Their masses are
different in the intermediate phase and coincide on the two
density plateaux. We give formulas for their approach to the
critical boundaries and for their continuation away from the
real chemical-potential axis.

Finally, imaginary chemical potential provides a setting in which
the same static model can be simulated directly. The $SU(N)$ center
fixes the Roberge--Weiss period to $2\pi/N$ in the quark angle
\cite{RW}. To compare different values of $N$, we resolve this
period before taking the limit. The comparison with $SU(3)$ is
performed at the same bare parameters and with the same static
determinant. It therefore tests separately the effect of a finite
number of colors and the homogeneous mean-field approximation.

The paper has four main sections. Section~\ref{sec:fixed-sector}
defines the matrix model and derives its resolvent.
Section~\ref{sec:mf} gives the mean-field equations, the three
phases and their critical boundaries, followed by explicit
limiting formulas and thermodynamic profiles.
Section~\ref{sec:screening} treats the two screening masses.
Section~\ref{sec:rw} discusses imaginary chemical potential and
the comparison with spatial $SU(3)$ simulations.
The appendices contain the general algebraic elimination,
the projective transformations, the source-Hessian calculation
and the simulation details.

\section{Static determinant and the one-site matrix model}\label{sec:fixed-sector}

Related unitary matrix potentials were studied by Russo and
collaborators \cite{Russo,RussoTierz,SantilliTierz}, and more recently by
Malik and by Malik and Ahmed \cite{Malik,MalikAhmed}. The latter papers
analyze the spectral and phase structure of closely related one-matrix
problems, including complex potentials and Fisher--Hartwig
singularities. Here the two sources are treated independently at the
one-site level and then determined by the spatial mean field. The
resulting density thresholds, spatial screening poles and comparison
with the same-action $SU(3)$ simulations require that second step.

On the spatial lattice the partition function is
\begin{align}
 Z_\Lambda={}&\int\prod_{\mathbf x}\dd U_{\mathbf x}\,
 \exp\left[\beta\sum_{\langle\mathbf x\mathbf y\rangle}
 \operatorname{Re}(\Tr U_{\mathbf x}\Tr U_{\mathbf y}^{\dagger})\right]\notag\\
 &\times \prod_{\mathbf x}e^{mNN_f}
 \left\{\det(1+h e^\mu U_{\mathbf x})
       \det(1+h e^{-\mu}U_{\mathbf x}^{\dagger})\right\}^{N_f}.
 \label{eq:spatial-action}
\end{align}
Each undirected bond is counted once, $U_{\mathbf x}\in SU(N)$, and
$m,\mu$ are dimensionless thermal parameters. We use $h=e^{-m}$,
$b=\beta d$ and $\kappa=N_f/N$. Throughout, the thermodynamic
logarithm is normalized per site and per $NN_f$, with its sign
chosen as $+\log Z_\Lambda$. This is the static effective spin
model, without spatial quark hopping; related effective and dual
formulations are given in Refs.~\cite{BCVdual,Fromm,DualPhase}.

We keep the notation $g_\pm$ for the shifted sources used inside
the matrix integral. Their relation to the physical sources is
$g_\pm=e^{\mp\mu}g_\pm^{\rm phys}$. With the present orientation
convention $g_+$ multiplies $U$ and $g_-$ multiplies $U^\dagger$.
The original loop averages are denoted by
$\mathrm w=\langle\Tr U/N\rangle$ and
$\mathrm w^\dagger=\langle\Tr U^\dagger/N\rangle$.
At real chemical potential the superscript $\dagger$ labels the
opposite insertion; the two averages need not be complex conjugates.

Let $q$ be the integer sector of the determinant constraint, and set $u=q/N_f$ and $v=q/N=\kappa u$.
The sources $\gp$ and $\gm$ and the determinant sector $q$ are independent
throughout this section. After shifting the fugacity factors into the sources and the sector weight,
and omitting the static scalar factor, the
one-site integral and its planar logarithm are
\begin{align}
 Z_q(\gp,\gm;h)&=\int_{U(N)}\dd U\, (\det U)^{-q}
 e^{N\Tr(\gp U+\gm U^{-1})}
 \bigl[\det(1+hU)\det(1+hU^{-1})\bigr]^{N_f},\\
 S_q(\gp,\gm;h,u)&=\lim_{\substack{N,N_f\to\infty\\N_f/N=\kappa}}
                       \frac{\log Z_q(\gp,\gm;h)}{NN_f}.
\end{align}
The limit holds $u=q/N_f$ fixed. The sign in the determinant
insertion fixes the sign of $v=\kappa u$. In the resolvent
calculation it is convenient to use the auxiliary function
$\Psi(\gp,\gm;h,v)=\kappa S_q(\gp,\gm;h,v/\kappa)$: its source
derivatives give the traces normalized by $N$. We retain this
auxiliary notation in the algebraic equations and Hessians.

The Fourier representation of the constraint $\det U=1$ inserts
$\sum_{q\in\mathbb Z}(\det U)^{-q}$ in the $U(N)$ integral.
If the quark fugacity is removed from the local determinant by
$z=e^\mu z_{\rm original}$, the constraint contributes
$e^{Nq\mu}$ to the $q$th term. Thus $q$ counts baryon charge in
the sector sum, whereas $Nq$ couples to the dimensionless
\emph{quark} chemical potential used here. At zero sources
the static determinant bounds the baryon charge by $|q|\le N_f$.
The exponential sources at a general off-shell point need not
obey this bound; the charge plateaux discussed below concern the
self-consistent static model.
Define
$M_+=\partial_{\gp}\Psi$, $M_-=\partial_{\gm}\Psi$ and
\begin{equation}
 \Gamma=\gp M_++\gm M_-,
 \qquad G=\frac h2\partial_h\Psi .
\end{equation}
We use the saddle continuously connected to the determinant-only
solution at zero spatial coupling, following the signed endpoint
conditions through the gap opening. Competing stationary solutions
must be compared by their free energies; algebraic roots alone do
not establish their dominance. This prescription and its possible
limitations apply to the thermodynamic and screening calculations
below.

The order of limits is important in using these definitions. At
real chemical potential, a filling of order one means $q$ is
of order $N_f$, and the local saddle changes when $v$ changes.
For the imaginary-potential comparison a center period instead
corresponds to $\operatorname{Im}\mu$ of order $1/N$. Terms with
$q=O(1)$ then have distinct Fourier phases even if all their
values of $v$ tend to zero. Thus the same fixed-sector integral
will be used in two different scalings. Replacing the sector
sum by its real-density extremum is appropriate for the
leading real-chemical-potential saddle; it would remove the
information about adjacent center periods in the imaginary
problem.

\subsection{Independent sources and the resolvent}

Let $\dd\rho$ be the normalized planar eigenvalue measure on
the selected contour and let
$\omega(z)=\int\dd\rho(\zeta)/(z-\zeta)$. Moving the unitary
Vandermonde to a holomorphic contour changes each eigenvalue
weight by a power of $z$. In particular, the double product in
the determinant can be written in terms of $z+h$ and
$z+h^{-1}$ with one inverse power of $z$ per flavor. Together
with the Vandermonde and $\det U^{-q}$ this gives the coefficient
$e=1+\kappa+v$ at the origin in $V'$. Across a regular cut,
the eigenvalue saddle obeys $\omega_++\omega_-=V'$.
Consequently $y=V'-2\omega$ changes sign across the cut.
The ratio $y/\mathcal R$ is then single-valued on the plane
and its only poles are the ones inherited from $V'$. This
is why the cut calculation reduces to matching a few residues
rather than solving an integral equation for each value of the
sources.

The effective force is
\begin{equation}
 V'(z)=-\gp+\frac{\gm}{z^2}+\frac{e}{z}
        -\frac{\kappa}{z+h}-\frac{\kappa}{z+h^{-1}},
 \qquad e=1+\kappa+v,\qquad d_{\mathrm p}=1+\kappa-v.
\end{equation}
Write $\mathcal R(z)=\sqrt{z^2-2Lz+p^2}\sim z$ at infinity, with signed values
$p=\mathcal R(0)$, $r=\mathcal R(-h)$ and $s=\mathcal R(-h^{-1})$.
The one-cut solution $y=V'-2\omega$ has the rational prefactor
\begin{equation}\label{eq:y}
 y(z)=\mathcal R(z)\left[-\frac{\gp}{z}+\frac{\gm}{pz^2}
  +\frac{\kappa h}{r\,z(z+h)}
  +\frac{\kappa}{hs\,z(z+h^{-1})}\right].
\end{equation}
The form follows by matching the leading term at infinity, the double
pole at zero, and the residues at the two determinant poles. Matching
the $1/z$ term at infinity, using $\omega(z)\sim1/z$, and the simple
pole at zero gives
\begin{align}
 \gp L+\gm/p+d_{\mathrm p}+\kappa(h/r+1/(hs))&=0,\\
 p\gp+\gm L/p^2+e-\kappa p(1/r+1/s)&=0.\\[-1mm]
 r^2=p^2+2hL+h^2,\\
 s^2=p^2+2L/h+h^{-2}.
\end{align}
Eliminating $L$ between the last two equations leaves the diagonal
quadric
\begin{equation}\label{eq:quadric}
 { R^2-h^2 S^2+(h^2-1)(P^2-Q^2)=0,\qquad
 [P:Q:R:S]=[p:1:r:s]. }
\end{equation}
This is the projective form of the signed endpoint data. The two
remaining pole conditions cut out the planar solutions on it.

The signs of $p,r,s$ cannot be chosen independently by taking
positive square roots. Normalization selects the sheet at infinity,
and continuation of that sheet fixes the signs at all three points.
This matters especially when a cut approaches one of the
determinant poles: a solution of the squared endpoint equations
without the signed pole conditions can describe a different saddle.
Once the signs have been fixed, the Laurent expansion of $\omega$
at infinity gives $M_+$ and its value at the origin gives $-M_-$.
The mass response instead uses the finite parts at the determinant
poles, as explained in Appendix~\ref{app:elimination}.

\subsection{Local observables and the neutral phase}

The endpoint equations determine the local observables without
requiring an explicit expression for the free energy. The dependence
on the mass and the two sources is algebraic at the level of its
first derivatives. The general elimination and the smaller equations
on special slices are given in Appendix~\ref{app:elimination}.
For the thermodynamic calculation it is simpler to solve the
endpoint equations themselves and integrate their first moments.

The three responses have different uses. The radial derivative
$\Gamma$ changes the overall source scale and so can reconstruct
the source dependence of $\Psi$ by integration from a reference
point. The separate $M_\pm$ enter two different self-consistency
conditions and cannot in general be replaced by $\Gamma$ alone.
Finally, $G$ gives the derivative with respect to the mass at
fixed sources. In particular, $G$ is constant on the massless
slice, even though the source response and the mass susceptibility
are nontrivial there. It is the derivative, rather than $\Psi$
itself, that is algebraic.

One can see the role of independent sources explicitly on the
regular full-support neutral branch, for $0<h<1$.
The positive- and negative-frequency couplings in the logarithm
of the local weight have coefficients
\[
 a_n=\gp\delta_{n1}+\kappa\,\frac{(-1)^{n+1}h^n}{n},
 \qquad
 b_n=\gm\delta_{n1}+\kappa\,\frac{(-1)^{n+1}h^n}{n}.
\]
For a saddle supported on the full circle or its regular contour
deformation the loop equation is diagonal in Fourier modes.
The planar logarithm, with the additive convention of the
fixed-sector integral, is $\sum_{n\ge1}n\,a_nb_n$. Therefore
\begin{equation}\label{eq:neutral-illustration}
 \Psi_{\rm fs}=\gp\gm+\kappa h(\gp+\gm)-\kappa^2\log(1-h^2),
 \quad M_+=\gm+\kappa h,\qquad M_-=\gp+\kappa h.
\end{equation}
At real symmetric sources the full-circle density first reaches
zero at $z=-1$. Continuing the full-support solution beyond
that point would give an unphysical eigenvalue density. The
gapped solution must then be computed with a cut and, at general
density, with the signed endpoint conditions. Equation
\eqref{eq:neutral-illustration} nevertheless gives a local check:
the two diagonal source curvatures vanish, while the mixed one
is unity. This is precisely the response used later on the
zero-density plateau.

\subsection{Vanishing fugacity: \texorpdfstring{$h=0$}{h=0} and the GWW model}

Here the determinant factors become unity while the sector insertion
remains. At finite $N$, for positive $\gp,\gm$, the change of
variable $z=\sqrt{\gm/\gp}\,w$ in the eigenvalue contour integral gives
\begin{equation}
 Z_q(\gp,\gm;0)=\left(\frac{\gp}{\gm}\right)^{Nq/2}
 Z_q^{\rm GWW}(H),\qquad H=2\sqrt{\gp\gm}.
\end{equation}
The GWW source \cite{GrossWitten,Wadia} is $N(H/2)\Tr(U+U^{-1})$ in this formula.
Consequently the radial source derivative is twice the usual
GWW generating function $g_{\rm GWW}=(H/2)\partial_H\Psi_{\rm GWW}$,
and $\gp M_+-\gm M_-=v$. This is an exact reduction, with the source
ratio entering only through the sector prefactor. It is distinct
from the many-flavor scaling that also produces GWW.

If exactly one source vanishes at $h=0$, the surviving sectors
are selected by the global rotation $U\mapsto e^{i\theta}U$.
For $\gp\ne0$, $\gm=0$, only $q\ge0$ contribute and every nonzero
sector has $Z_q\propto \gp^{Nq}$. For $\gp=0$, $\gm\ne0$, the
corresponding condition is $q\le0$ and $Z_q\propto \gm^{-Nq}$.
Consequently $\Gamma=|v|$ in the surviving sectors (and $Z_0=1$).
The massless one-source quintic has no nontrivial counterpart
in this exact zero-fugacity, one-source slice.

\subsection{The quenched limit at fixed sources}\label{sec:quenched-site}

For fixed $h$ and fixed $u=q/N_f$, taking $\kappa\to0$ sends
$v=\kappa u$ to zero. The determinant and the sector insertion cease
to contribute at order $N^2$. Keep $\gp,\gm>0$ fixed and write
$g=\sqrt{\gp\gm}$. Rescaling the eigenvalue contour reduces the remaining
neutral integral to the ordinary GWW problem with source
$Ng\Tr(U+U^{-1})$. In the $N^2$ normalization its logarithm is
\begin{equation}\label{eq:quenched-gww}
 \Psi_{\rm YM}(g)=
 \begin{cases}
 g^2,& 0\le g\le\tfrac12,\\
 2g-\tfrac12\log(2g)-\tfrac34,& g\ge\tfrac12 .
 \end{cases}
\end{equation}
The first branch follows from the full-circle density
$1+2g\cos\theta$; on the arc branch the GWW first moment is
$1-1/(4g)$. Integrating twice that moment and imposing continuity at
$g=1/2$ gives the second branch of \eqref{eq:quenched-gww}.
The source asymmetry is retained in the individual moments:
\begin{equation}\label{eq:quenched-moments}
 M_+=\frac{\gm}{g}\mathrm w(g),\qquad
 M_-=\frac{\gp}{g}\mathrm w(g),\qquad
 \mathrm w(g)=\frac12\Psi'_{\rm YM}(g)=
 \begin{cases}
 g,&g\le\tfrac12,\\
 1-\dfrac{1}{4g},&g\ge\tfrac12 .
 \end{cases}
\end{equation}
Thus $\Gamma=2g\mathrm w(g)$, whereas $G$ vanishes at leading order.
The limiting expressions at $\gp\gm=0$ are taken from the first branch,
$\Psi_{\rm YM}=\gp\gm$. If one instead holds $v$ nonzero as
$\kappa\to0$, one keeps the determinant insertion even though the
fermion determinants disappear, and recovers the \emph{fixed-sector}
GWW integral of the preceding subsection. These are different limits.
At fixed nonzero sources the one-site action $S_q$ contains
$\Psi_{\rm YM}/\kappa$ and is generally singular in the quenched
limit. The finite GWW function here describes its pure-glue
coefficient. The behavior after imposing spatial
self-consistency is different; see Section~\ref{sec:quenched-mf}.

\section{Thermodynamics at real chemical potential}
\label{sec:mf}

We now embed the local solution in the Polyakov loop model.
For $m>0$ the equilibrium filling can remain zero over a finite
interval of chemical potential. An intermediate phase follows,
in which baryons enter continuously. Its upper boundary is
saturation, where the filling is one and the free energy becomes
linear. Negative chemical potential gives the charge-conjugate
picture. We consider $0\le b<1$ and determine both boundaries
from the endpoint slopes of the free energy.

The chemical potential enters the sector sum with weight
$e^{Nq\mu}$. Let $g_+^{\rm phys},g_-^{\rm phys}$ couple to the original
loop orientations. The contour change
$z=e^\mu z_{\rm original}$ gives
$\gp=e^{-\mu}g_+^{\rm phys}$ and
$\gm=e^\mu g_-^{\rm phys}$. Thus the two determinant
factors have the common fugacity $h$, and the interaction
counterterm is unchanged since $g_+^{\rm phys}g_-^{\rm phys}=\gp\gm$. In units of $NN_f$
per site, the logarithm of the saddle weight is
\begin{align}
 S_{\rm eff}[\kappa,\gp,\gm,h;u]
   &=m+S_q(\gp,\gm;h,u)+\mu u,\label{eq:site-effective}\\
 f(\gp,\gm,u;\mu)&=S_{\rm eff}[\kappa,\gp,\gm,h;u]
                       -\frac{\gp\gm}{\kappa b}.
 \label{eq:mf-action}
\end{align}
The stationary value $f_*=f_{SU(N)}$ is the logarithmic free energy
per site and per $NN_f$. The notation $S_{\rm eff}$ refers to the
one-site action before subtracting the mean-field counterterm.
At $b=0$ one sets $\gp=\gm=0$ and omits the counterterm.

For the normalization in Eq.~\eqref{eq:spatial-action}, varying
the interaction with respect to the loop at one site gives
$\beta/2$ times the sum of the opposite loop orientations on
its $2d$ neighbors. Thus a homogeneous saddle has
$g_+^{\rm phys}=b\,\mathrm w^\dagger$ and
$g_-^{\rm phys}=b\,\mathrm w$, where $\mathrm w$ and $\mathrm w^\dagger$ denote
the two original normalized traces. They need not be equal
at real chemical potential. The two sources are varied
independently in the local calculation; setting them equal
before differentiation would eliminate precisely the response
that distinguishes the two orientations in the dense regime.

For each fixed $u$, solve the source equations on the continued
real mean-field branch,
\begin{equation}\label{eq:mf-sources}
 \gm=bM_+,\qquad \gp=bM_-.
\end{equation}
In terms of the effective action this is precisely the form used
in the original mean-field formulation:
\begin{equation}
 \frac{g_-}{\kappa b}=\partial_{g_+}S_{\rm eff},\qquad
 \frac{g_+}{\kappa b}=\partial_{g_-}S_{\rm eff},\qquad
 \partial_u S_{\rm eff}=0
 \quad\hbox{on an interior sector saddle}.
\end{equation}
The last condition is imposed at fixed sources. At $u=\pm1$
it is replaced by the appropriate one-sided extremum condition.

Let $f_b(u)$ denote the resulting action without $\mu u$.
The thermodynamic prescription is to compare the admissible
stationary branches by their action and maximize $f_b(u)+\mu u$
in the sector variable. On an interior branch,
\begin{equation}\label{eq:mf-u}
 \mu=-f_b'(u)=-\partial_v\Psi(\gp,\gm;h,v)
     =\int_{\mathcal C}\log z\,\dd\rho(z),\qquad
 \frac{\dd f_*}{\dd\mu}=u_* .
\end{equation}
The logarithm follows the continued saddle contour. The source
derivatives in $f_b'$ cancel by \eqref{eq:mf-sources}.
In particular, \eqref{eq:mf-u} is a saddle equation for the
sector variable, not a further polynomial equation for the
one-site resolvent. A convenient way to reconstruct its reduced
action from the exact moments is
\begin{equation}\label{eq:mf-envelope}
 f_b(u)=f_0(u)+\frac1\kappa\int_0^b
 M_+(u,\widetilde b) M_-(u,\widetilde b)\,\dd\widetilde b .
\end{equation}
Here $f_0(u)$ includes the static normalization $m$ and is computed
from the source-free matrix model. Equations \eqref{eq:mf-sources}
and \eqref{eq:mf-u} give the interior density; the following
endpoint calculations do not require differentiating a numerical
solution for that density.

The coupling integral follows from the envelope property:
at stationary sources the only explicit $b$-derivative of
$f-\mu u$ is $\gp\gm/(\kappa b^2)$. Using
\eqref{eq:mf-sources} gives the integrand in
\eqref{eq:mf-envelope}. In the interior the graph of $-f_b'(u)$
can therefore be read as the chemical potential needed to
hold a given filling. The maximizer is $u=0$ or $u=1$ when the
corresponding one-sided derivative pins it at that endpoint.
The thresholds can be found from these endpoint slopes without
a finite-density interpolation. In our plots, the interpolation
is used for the observables \emph{between} the thresholds.

For the physical loop orientations the sources are shifted
back after solving the saddle. In particular, if the shifted
neutral solution has $\gp=\gm$ and $M_+=M_-$, this does not imply
that the two measured loops agree at $\mu\ne0$.
Likewise, the filling is determined by differentiation with
respect to the external chemical potential, not by holding the
original sources at unrelated values while varying $u$ and
$\mu$. These distinctions are important when differentiating
the local solution for the screening calculation. The mean-field
interaction gives a stationarity condition on the sources; the
charge extremum at fixed coupling is a separate step.

\subsection{The three density phases and their observables}\label{sec:mf-observables}

We characterize the phases by the filling and the mass response.
This also specifies their meaning at nonzero quark mass, where the
condensate need not be a symmetry order parameter. Write $\mu_-$ and $\mu_+$ for the onset and saturation
chemical potentials, and set $x=h^2$ and $J=b\kappa/(1-b)$
in the plateau formulas. The thresholds separate three regimes
when $m>0$ and the positive onset threshold exists. Table~\ref{tab:mf-phases} describes them for
$\mu\ge0$ on the real mean-field branch considered above.
In the zero-density phase the free energy and condensate have a
plateau, but the two Polyakov loops have opposite exponential
dependences on $\mu$. Between the two thresholds the free energy
has an increasing slope, $0<u_*<1$. A positive $\mu$ can enter this
interior regime directly, without a zero-density interval, if
the onset condition fails. The massless cut is such an example.
In the saturated phase the free energy has unit slope and is
independent of $m$, but one of the two loops still responds to
$\mu$. Negative $\mu$ reverses the charge and exchanges the two loops.

\begin{table}[htbp]
\centering\small
\begin{tabular}{@{}lccc@{}}
\hline
 & Zero density & Partial filling & Saturation\\
\hline
Range & $0\le\mu<\mu_-$ & $\mu_-<\mu<\mu_+$ & $\mu>\mu_+$\\
$u_*$ & $0$ & $-f_b'(u_*)=\mu$ & $1$\\
$f_*$ & $m-\kappa\log(1-x)+Jx$ &
$ f_b(u_*)+\mu u_* $ & $\mu$\\[1mm]
$(\mathrm w,\mathrm w^\dagger)$ & $\dfrac{\kappa h}{1-b}(e^{-\mu},e^\mu)$ &
$(e^{-\mu}M_+,e^\mu M_-)$ &
$\left(\dfrac{\kappa(h+h^{-1})}{1-b}e^{-\mu},\,0\right)$\\[2mm]
$\Sigma$ & $1-\dfrac{2\kappa x}{1-x}-2Jx$ &
$1-2G/\kappa$ & $0$\\
\hline
\end{tabular}
\caption{The three phases on the continued real saddle. The interior
formula for the condensate applies also at a regular endpoint when
evaluated on the appropriate branch. If there is no lower line, the
first column has no interval and the interior regime starts at
$\mu=0^+$.}
\label{tab:mf-phases}
\end{table}

Here $\mathrm w=\langle\Tr U/N\rangle$ and
$\mathrm w^\dagger=\langle\Tr U^{-1}/N\rangle$ refer to the original,
unshifted matrix variable. Restoring the chemical-potential factors,
\begin{equation}\label{eq:mf-observables}
 \mathrm w=e^{-\mu}M_+,\qquad \mathrm w^\dagger=e^{\mu}M_-,\qquad
 n_B=\partial_\mu f_*=u_*,\qquad
 \Sigma=\partial_m f_*=1-\frac{2G}{\kappa}.
\end{equation}
The normalized baryon density $n_B=\langle B_{\rm tot}\rangle/(VN_f)$
equals the filling $u_*$ used in the plots, where $V$ is the number
of spatial sites. The derivative with respect to the quark chemical
potential gives this normalization because $\mu$ couples to
$N B_{\rm tot}$. The saturation values are therefore $\pm1$
at every $\kappa$. We plot $f_*$
with the sign of $\log Z$ as in \eqref{eq:mf-action}; a convention
with minus the logarithm reverses its sign and the sign of its mass
derivative. Mass derivatives hold $\mu,b,\kappa$ fixed; chemical-potential
derivatives hold $m,b,\kappa$ fixed. The
source and interior sector derivatives drop out by stationarity;
on an endpoint of the sector interval, $u$ is held at that endpoint.

\subsection{A common coordinate for onset and saturation}

Write
\begin{equation}\label{eq:mf-coordinate}
 x=h^2,\qquad J=\frac{b\kappa}{1-b},\qquad
 H=J(1+x),\qquad w=-hz,\qquad D(w)=(w-1)(w-x).
\end{equation}
The two determinant poles are at $w=x$ and $w=1$. We use $\mu>0$;
the negative half of the phase diagram is obtained by charge
conjugation.

\paragraph{Lower line: departure from zero density.}
On the ungapped $u=0$ branch, $\gp=\gm=g=Jh$ by
\eqref{eq:mf-sources}, and the reduced endpoint action is
$f_b(0)=m-\kappa\log(1-x)+Jx$. The density on the circle and its analytic
continuation to $z=-w/h=-e^\eta$ are
\begin{align}
 \rho(z)&=1+g(z+z^{-1})
   +\kappa\left(\frac{hz}{1+hz}+\frac{h}{z+h}\right),\\
 \rho_\eta&=1-J(w+x/w)
     -\kappa\left(\frac{w}{1-w}+\frac{x}{w-x}\right),
     \qquad w=he^\eta .
\end{align}
On the outside sheet of the full-support solution,
$V'-2\omega=-\rho(z)/z$. The incipient cut is at the first zero
of this spectral differential. The one-sided sector slope, or
cost per unit charge for the first baryon, is its integral from
$z=-1$ to the zero, with sign chosen for positive $\mu$:
$\mu_-=-f_b'(0^+)=\int_0^{\eta_-}\rho_\eta\,\dd\eta$.
Hence the onset equation and chemical potential are
\begin{align}
 {P_-(w)=[w-J(w^2+x)]D(w)
      +\kappa w(w^2-2xw+x)=0,\quad h<w_-<1,}
      \label{eq:mf-lower-quartic}\\
 {\mu_-=m+\log w_- -J(w_--x/w_-)
       +\kappa\log\frac{1-w_-}{1-x/w_-}.}
      \label{eq:mf-mu-lower}
\end{align}
The root exists only if $\rho_{\eta=0}=1-2Jh-2\kappa h/(1-h)>0$.
In that case $\rho_\eta$ is strictly decreasing for $0<\eta<m$,
and tends to $-\infty$ as $\eta\to m$. Thus the interval in
\eqref{eq:mf-lower-quartic} selects exactly one root.
Equality in the existence condition is the junction with the
$\mu=0$ gap-opening line, where $w_-=h$ and $\mu_-=0$.
Beyond it there is no positive onset threshold; in particular
there is none at $h=1$.

For computation the reciprocity $w\leftrightarrow x/w$ further
reduces the quartic. Set $Y=w+x/w$, $K=1+\kappa+J(1+x)$,
and $C_0=1+(1+2\kappa)x$. Then
\begin{align}
 JY^2-KY+C_0&=0,\\
 Y_-&=\frac{2C_0}{K+\sqrt{K^2-4JC_0}},\qquad
 w_-=\frac{Y_-+\sqrt{Y_-^2-4x}}2 .
\end{align}
This prescription is continuous at $b=0$ and avoids cancellation
in the smaller quadratic root.

The sign of the integrated lower edge can be read directly
from the contour. With $z=-e^\eta$ one has $\dd z/z=\dd\eta$.
The neutral resolvent gives $y(z)\dd z=-\rho_\eta\dd\eta$,
so minus this integral is positive up to the zero at
$\eta_-$. The $1$ in $\rho_\eta$ integrates to $\eta_-$;
the first harmonic integrates to $-2g\sinh\eta_-$, and the
two determinant terms give the logarithm of a ratio in
\eqref{eq:mf-mu-lower}. Both determinant arguments are
positive on the selected interval. This fixes the real
branch of the logarithm and avoids a separate analytic
continuation of a fitted free-energy curve.

\paragraph{Upper line: arrival at saturation.}
On the $u=1$ branch, $M_-=\gp$ and
$M_+=\kappa(h+h^{-1})+\gm$; therefore
$\gp=0$ and $\gm=J(h+h^{-1})$. The reduced endpoint action is
$f_b(1)=0$. At these sources one can factor the maximum determinant
power from the fixed-sector integrand. The sector one unit below
saturation, relative to the saturated sector, is then governed
by the exact coefficient
\begin{equation}
 [t^N]\,(1-ht)^{-N_f}(1-t/h)^{-N_f}e^{NBt}.
\end{equation}
The contour saddle lies at $0<t<h$.
With $t=h/w$ (equivalently $z=-1/t$), its condition gives
\begin{align}
 {P_+(w)=(H-w)D(w)
       +\kappa w[(1+x)w-2x]=0,\quad w_+>1,}
       \label{eq:mf-upper-cubic}\\
 {\mu_+=-f_b'(1^-)=m+\log w_+
   -\kappa\log(1-x/w_+)-\kappa\log(1-1/w_+)+H/w_+.}
       \label{eq:mf-mu-upper}
\end{align}
In the inverse coordinate $t/h=1/w$, the saddle function is
strictly convex on $(0,1)$, so this root is unique. Retaining the
same clearing denominator $wD(w)$ as at onset gives the quartic
$wP_+(w)=0$, but $w=0$ is extraneous; the minimal upper equation in
this coordinate is cubic because $\gp=0$. At $h=1$, the poles
coalesce and the upper polynomial has a removable $w-1$ factor,
leaving $(H-w)(w-1)+2\kappa w=0$.

\subsection{Ordering and a density calculation}

\begin{figure}[tbp]
 \centering
 \includegraphics[width=\textwidth]{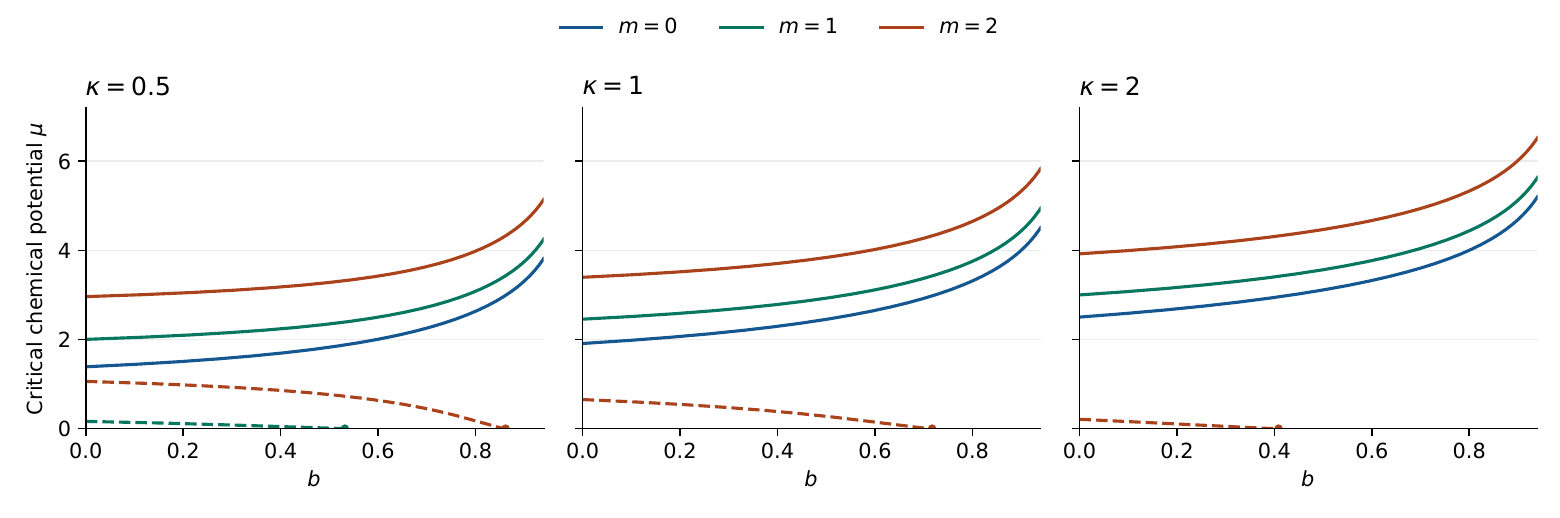}
 \caption{Onset and saturation as functions of the spatial coupling
$b$, for $\kappa=1/2,1,2$ from left to right. Colors mark $m=0,1,2$.
The dashed curves are $\mu_-$; the solid curves of the same color
are $\mu_+$. The interval between them has $0<u_*<1$.
The dots indicate that $\mu_-$ reaches zero; a missing dashed
curve denotes no zero-density interval at positive $\mu$.
For $m=0$ there is no positive onset line in any panel.}
\label{fig:crit}
\end{figure}

Figure~\ref{fig:crit} illustrates how the onset line can end
at $\mu=0$ while the saturation line moves to larger $\mu$.

\begin{figure}[tbp]
 \centering
 \includegraphics[width=.94\textwidth]{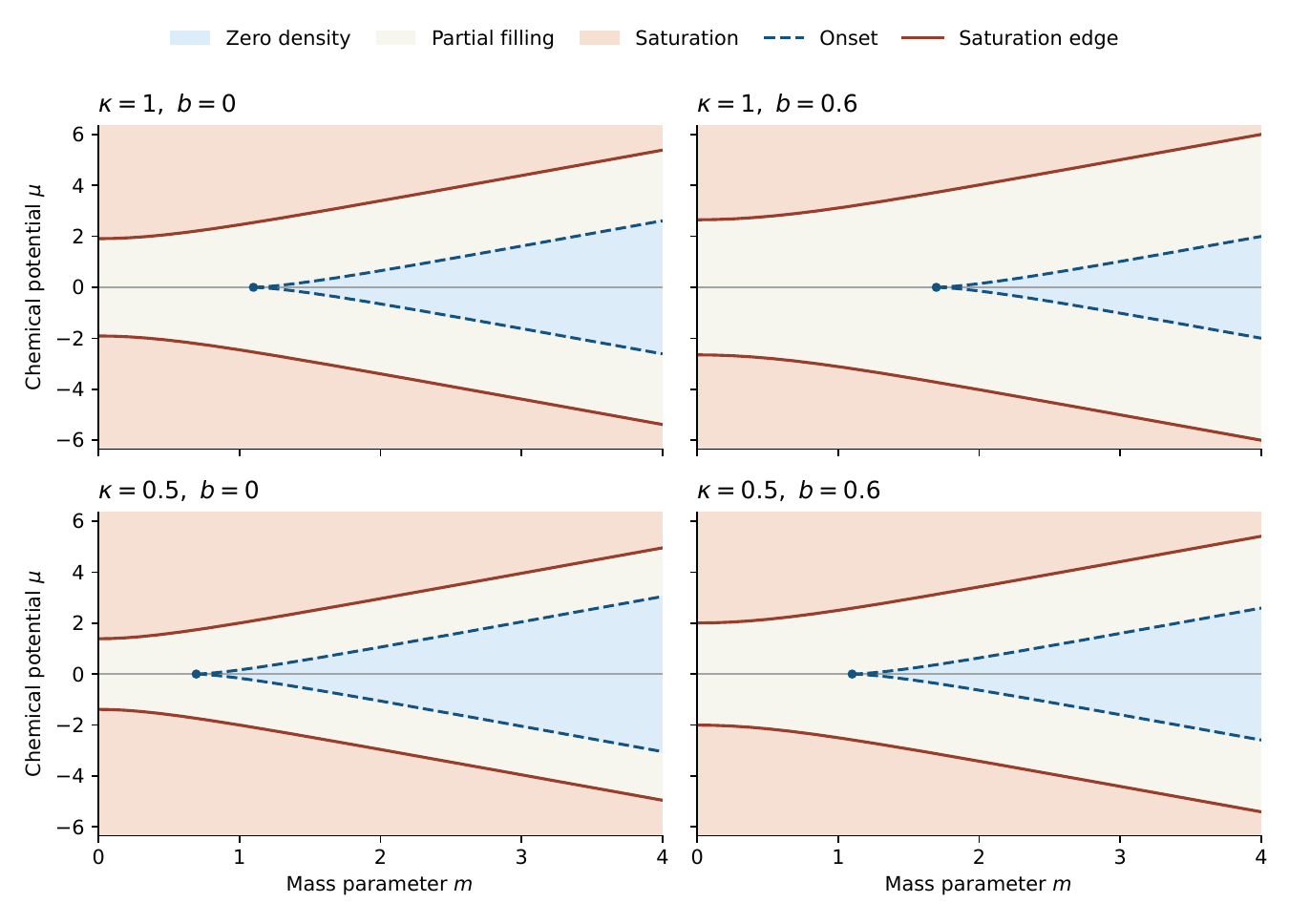}
 \caption{Real-density phase diagrams on the continued homogeneous
 branch in the $(m,\mu)$ plane. Top and bottom rows: $\kappa=1$ and
 $\kappa=1/2$. Left and right: $b=0$ and $b=0.6$. The dashed edges
 are $\pm\mu_-(m)$, and the solid edges are $\pm\mu_+(m)$.
 Blue shading marks $u_*=0$; the light background between
 the shaded regions denotes partial filling, $0<|u_*|<1$
 away from $\mu=0$. The outer shading denotes saturation,
 $u_*=\operatorname{sgn}\mu$. Each blue dot marks the first
 appearance of a positive onset threshold as $m$ is increased.}
 \label{fig:phase-maps}
\end{figure}

Figure~\ref{fig:phase-maps} gives the complementary phase
maps in $(m,\mu)$ for $\kappa=1$ and $1/2$. Both signs of the
chemical potential are shown. At fixed $b$, the zero-density
region first opens at the neutral gap threshold $m_{\rm gap}$.
For example, at $b=0$ it opens at $m_{\rm gap}=\log(1+2\kappa)$.
Raising $b$ to $0.6$ moves this opening from $1.099$ to $1.696$
at $\kappa=1$, and from $0.693$ to $1.099$ at $\kappa=1/2$.
There is no phase boundary along $\mu=0$ to the left of this
opening. On the negative-$\mu$ side the saturated phase has
$u_*=-1$; on the positive side it has $u_*=1$.

When the lower line exists, the positivity of $\rho_\eta$ before
its first zero and the signs of the terms in
\eqref{eq:mf-mu-lower}--\eqref{eq:mf-mu-upper} imply
\[
              0<\mu_-<\eta_-<m<\mu_+.
\]
The two thresholds of the continued branch therefore cannot
intersect at finite $\kappa>0$. By differentiating at their
stationary points one also finds
\begin{equation}
 \partial_b\mu_-=-\frac{\kappa(w_--x/w_-)}{(1-b)^2}<0,\qquad
 \partial_b\mu_+=\frac{\kappa(1+x)}{(1-b)^2w_+}>0.
\end{equation}
The onset line ends at the gap-opening condition rather than
crossing the upper line.

The reciprocal pair of lower roots has a simple interpretation:
$w\mapsto h^2/w$ changes $\eta$ to $-\eta$ and leaves the
analytic continuation of the neutral density invariant. The
positive-$\mu$ onset uses $\eta>0$ only. The upper saddle, by
contrast, is most readily understood in the fugacity variable
$t=h/w$. The coefficient integral contains $t^{-N}$, whose
contribution to the exponent per color is $-\log t$.
Adding the two logarithms from the inverse determinant and
the linear term $Bt$ gives exactly \eqref{eq:mf-mu-upper}.
Its derivative selects a root with $0<t<h$; zeros reached by
clearing denominators across either pole have no physical role.
These restrictions explain why the full algebraic equations
can have extra roots although the two thresholds are unique
on the branch at hand.

At fixed $m$ the onset condition also gives a simple coupling
criterion. Write $\rho_0=1-2\kappa h/(1-h)$ for the value at
$z=-1$ with zero spatial coupling. If $\rho_0>0$, the positive
onset interval ends at
\[
 b_{\rm gap}=\frac{\rho_0}{\rho_0+2\kappa h}.
\]
When $\rho_0\le0$, there is no positive onset interval even
at $b=0$. The upper line behaves differently near $b=1$ at
fixed $m,\kappa>0$: in the cubic, $H\to\infty$ and $w_+\sim H$.
Then $\mu_+=m+\log H+1+o(1)$. The approach to that coupling
therefore cannot be interpreted as an intersection of the two
thresholds at finite $\mu$. This statement is a limit from
below and does not solve the saddle at $b=1$.

For example, at $\kappa=1$ and $m=2$ the exact thresholds and the
interior solutions of \eqref{eq:mf-u} are
\begin{center}
 \begin{tabular}{c|rrrr}
 $b$ & $\mu_-$ & $\mu_+$ & $u_*(\mu=2)$ & $u_*(\mu=3)$\\ \hline
 $0$   & $0.651401$ & $3.395473$ & $0.492491$ & $0.903361$\\
 $0.3$ & $0.471409$ & $3.600854$ & $0.491079$ & $0.860704$\\
 $0.6$ & $0.144630$ & $4.019266$ & $0.489149$ & $0.807783$
 \end{tabular}
\end{center}
For the interior entries we continued the exact one-cut pole
equations together with \eqref{eq:mf-sources}, integrated
\eqref{eq:mf-envelope}, and inverted $-f_b'$. Doubling the
continuation grid changed the quoted densities by less than
$6\times10^{-8}$. The conjugate potential was increasing on these
numerical cuts. The analytic endpoints locate the appearance of density and
saturation without an interpolation in the interior.

\subsection{Free energy, Polyakov loops and chiral response}

The phase boundaries identify the two endpoints of the partially
filled regime. Its interior is resolved by the sector equation
\eqref{eq:mf-u}. The free energy, density, loop orientations and
mass derivative can thus be plotted on the same chemical-potential
axis, with the plateau values fixed analytically.

For evaluating the last expression it is useful to restore the
static factor directly in the integrand:
$e^m(1+hz)(1+h/z)=2\cosh m+z+z^{-1}$.
Writing $W(z)=2z\omega(z)-1$ in terms of the resolvent of
Section~\ref{sec:fixed-sector}, the resulting mass derivative is
\begin{equation}\label{eq:mf-sigma-res}
 \Sigma=\int_{\mathcal C}\frac{2\sinh m}{2\cosh m+z+z^{-1}}\,\dd\rho(z)
 =\frac{W(-1/h)-W(-h)}2 .
\end{equation}
The apparent poles in $W$ at the two determinant zeros cancel
when the pole conditions are enforced. On the regular $m=0$ saddle
the first mass derivative vanishes; the coalescing poles give the
nontrivial response $\chi_m=\partial_m\Sigma|_{m=0}=-W'(-1)$.
In saturation both responses vanish (away from the critical point).
The same evenness in $m$ removes the linear mass dependence of the
stationary sources and sector on the regular massless branch.

\begin{figure}[tbp]
 \centering
 \includegraphics[width=.83\textwidth]{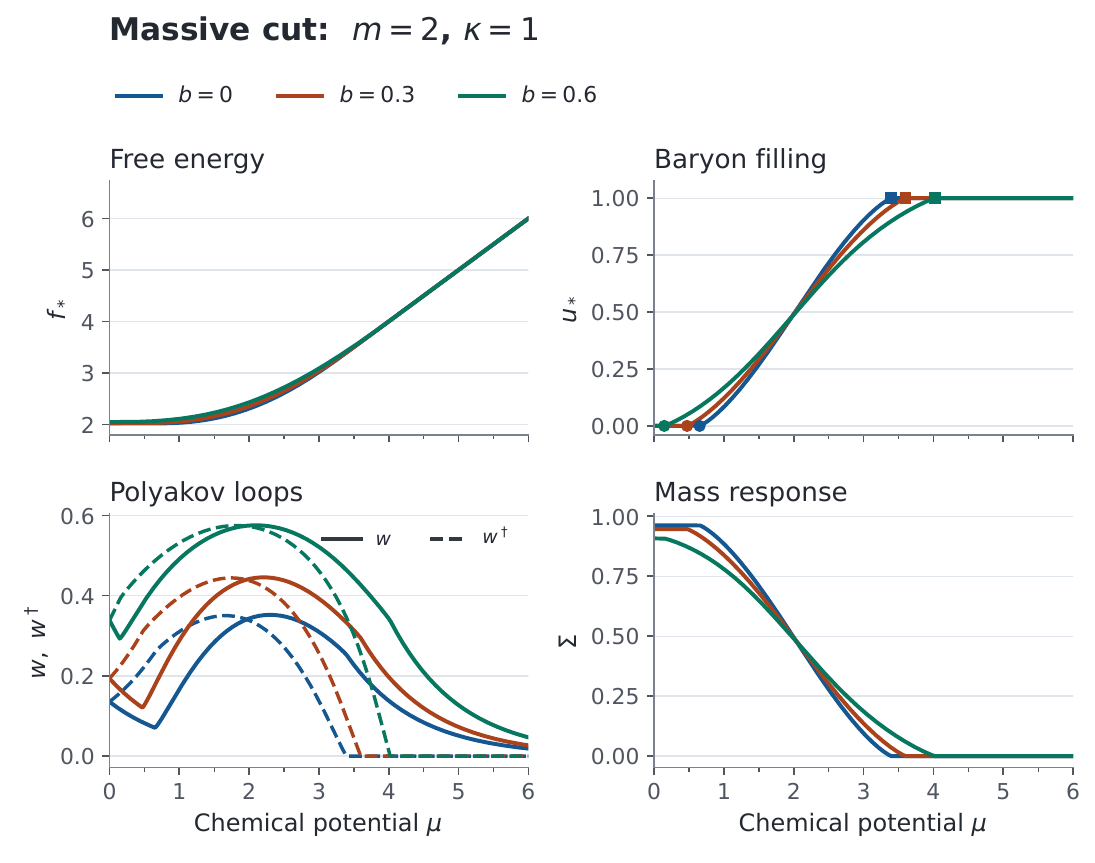}
 \caption{Real-density observables at $m=2$, $\kappa=1$,
for $b=0,0.3,0.6$. Top left: logarithmic free energy in $NN_f$
units per site, with $\partial_\mu f_*=u_*$.
Top right: filling, with the onset and saturation marked by circles
and squares. Bottom left: original loop $\mathrm w$ (solid) and
$\mathrm w^\dagger$ (dashed). Bottom right: $\Sigma=\partial_m f_*$,
which vanishes at saturation. For a given color the free-energy
plateau, change in the loop behavior and condensate drop can be
read against the thresholds in the filling panel.}
\label{fig:mf-massive}
\end{figure}

\begin{figure}[tbp]
 \centering
 \includegraphics[width=.83\textwidth]{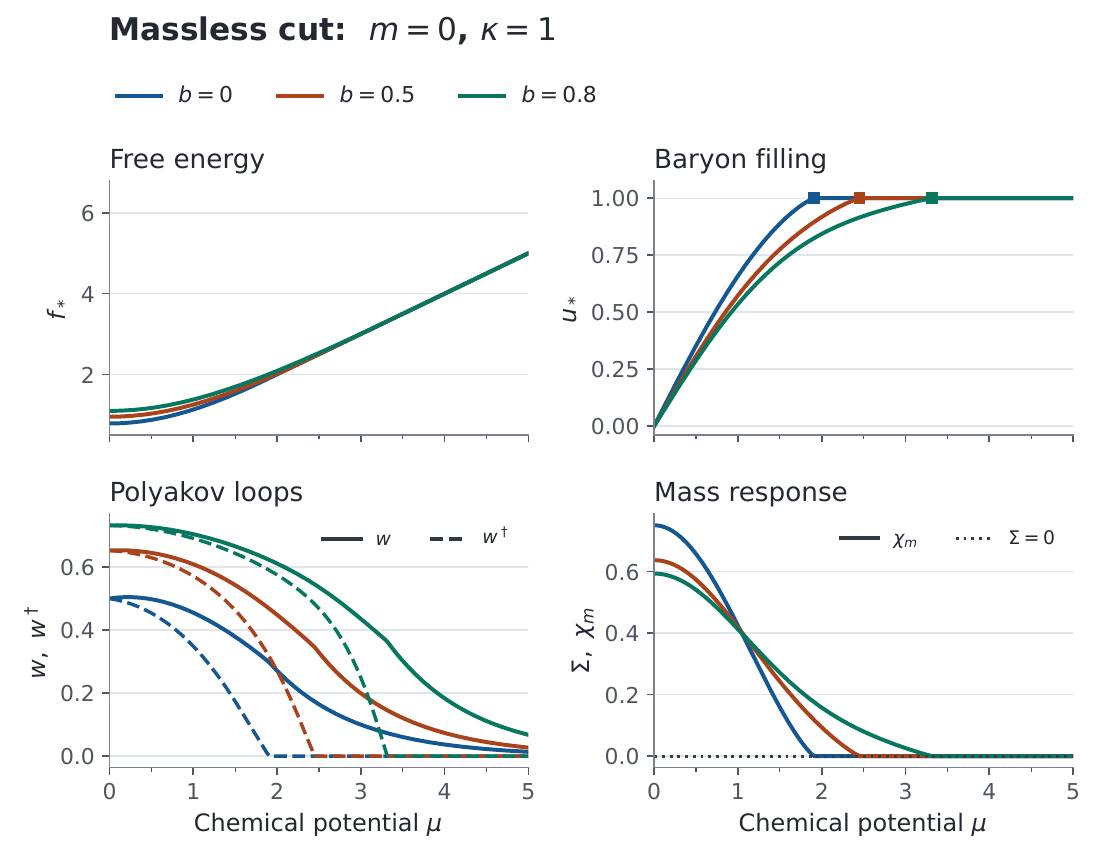}
 \caption{Massless cut at $\kappa=1$
for $b=0,0.5,0.8$. The free energy and filling (top) have
no positive-$\mu$ zero-density interval; squares mark saturation.
The two loop orientations are solid and dashed in the lower left.
In the lower right the common dotted zero line is the condensate,
while the colored curves are $\chi_m=\partial_m\Sigma|_{m=0}$.
The latter distinguishes the couplings even though $\Sigma$ itself
vanishes in the massless theory.}
\label{fig:mf-massless}
\end{figure}

\begin{figure}[tbp]
 \centering
 \includegraphics[width=.83\textwidth]{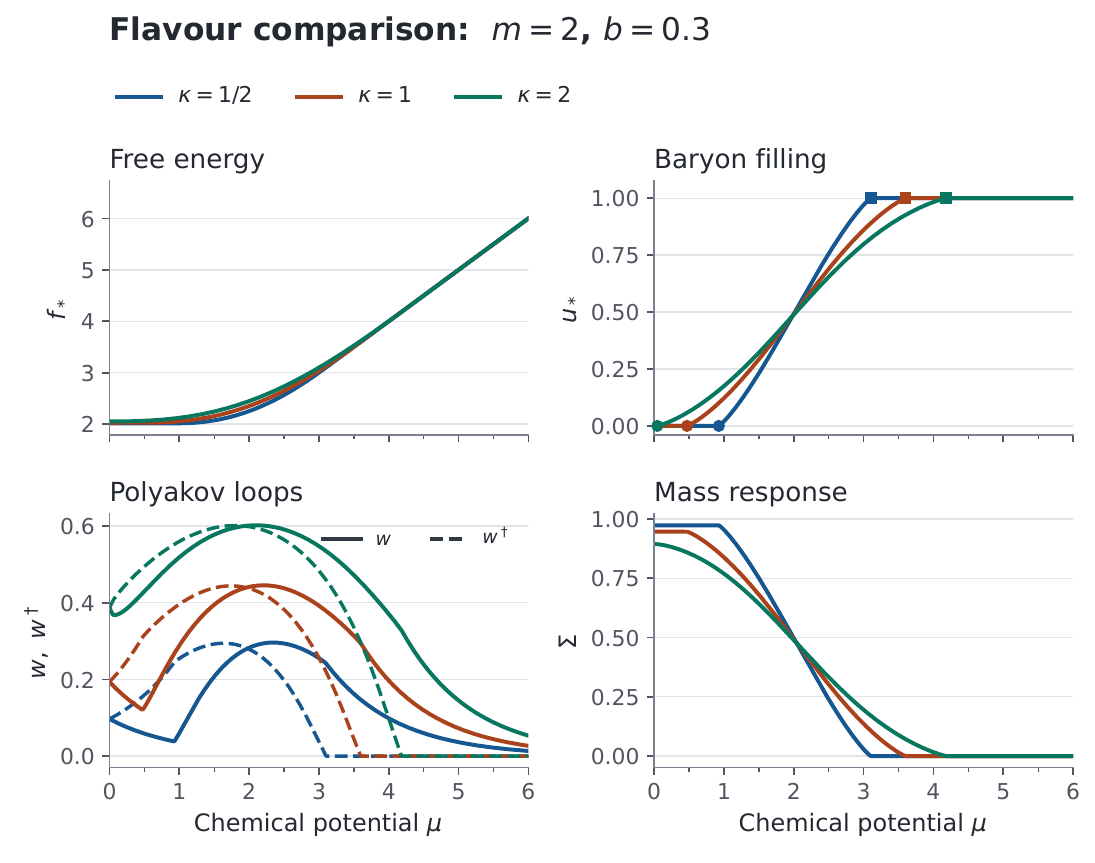}
 \caption{Flavor dependence at fixed $m=2$, $b=0.3$.
Colors denote $\kappa=1/2,1,2$. The panels show, clockwise from the
upper left, the free energy, the filling, the condensate, and the
two Polyakov loops (solid $\mathrm w$, dashed $\mathrm w^\dagger$). Filling is
normalized by $N_f$ at each $\kappa$, so all curves saturate at one.
Circles and squares in the filling panel locate $\mu_-$ and $\mu_+$,
respectively. Increasing $\kappa$ widens the interval of partial
filling on these cuts.}
\label{fig:mf-flavour}
\end{figure}

Figures~\ref{fig:mf-massive}--\ref{fig:mf-flavour} show the four
observables on matched cuts. The solid and dashed loop curves use
the \emph{same} sources; they differ by the choice of insertion
and by the factors in \eqref{eq:mf-observables}. The massive
$\kappa=1$ cut displays all three regimes; the massless cut isolates
the interior and saturated regimes. The flavour scan leaves $b$
and $m$ fixed. Along it, increasing $\kappa$ from $1/2$ to $2$
changes $(\mu_-,\mu_+)$ from $(0.927,3.102)$ to $(0.046,4.184)$;
the zero-density plateau nearly disappears while the upper threshold
moves to the right.
The curves use the pole equations and the source self-consistency
equations, with the sector action obtained from
\eqref{eq:mf-envelope}. We select the maximum over $0\le u\le1$
on the continued branch. On the plotted cuts the interpolated
reduced action is strictly concave. The analytic endpoints determine
the profiles on their plateaux; the nontrivial mass derivative was
also checked against a finite difference of the optimized free energy.


It is useful to distinguish the two ways in which a plateau
appears in the plots. The filling can be zero while the
condensate and the loops remain nonzero; in that interval the
$\mu$-dependence of the original loops arises from the factors
$e^{\mp\mu}$ relating them to the shifted sources. Saturation
instead fixes the baryon filling at its maximum and gives the
linear logarithmic free energy. The vanishing condensate
there is the mass derivative of that linear expression, not
the same statement as $\Sigma=0$ on the entire massless cut.
The mass susceptibility in Fig.~\ref{fig:mf-massless} can
remain nonzero even where the massless condensate is zero.

\FloatBarrier
\subsection{The massless model}\label{sec:massless-explicit}

The massless limit is a useful illustration of the difference
between the two critical boundaries. At $m=0$ the determinant
zeros coincide at $z=-1$. There is no interval of positive
chemical potential with zero density: the filling starts
continuously from $u=0$. Saturation still occurs at a finite
chemical potential for every finite $\kappa$ and $b<1$.

Putting $h=1$ in the upper equation removes one factor
$w-1$. With $J=b\kappa/(1-b)$ the remaining equation and its
physical root are
\begin{align}
 (2J-w)(w-1)+2\kappa w&=0,\\
 w_+&=\frac{1+2\kappa+2J+
       \sqrt{(1+2\kappa+2J)^2-8J}}{2}>1.
\end{align}
The massless saturation boundary is therefore explicit:
\begin{equation}\label{eq:massless-upper-explicit}
 {\mu_+(0,b,\kappa)=
  \log w_+-2\kappa\log(1-w_+^{-1})+\frac{2J}{w_+}.}
\end{equation}
In particular, the one-dimensional QCD limit gives
\begin{equation}
 \mu_+(0,0,\kappa)=
 (1+2\kappa)\log(1+2\kappa)-2\kappa\log(2\kappa).
\end{equation}
The negative boundary follows by charge conjugation.
This formula is continuous as $b\to0$ and is convenient
for comparing the massless curves without a numerical root search.

The small-density region can also be described analytically.
At $\mu=0$ the mean field is symmetric. Let
\[
 t_0=\sqrt{1-\sigma_0},\qquad
 \mathcal L_0=-\log(1-\sigma_0),
\]
where $\sigma_0$ is the opening parameter of the neutral cut,
defined in Eq.~\eqref{eq:screen-geometry}.
The pole condition and source equation give
\begin{equation}\label{eq:neutral-massless-quintic}
 {
 [(1+\kappa)t_0-\kappa](1-b+bt_0^4)
 -2b\kappa t_0(1-t_0)(1-t_0^2)=0.}
\end{equation}
The root is chosen in $0<t_0<1$ by continuation from
$t_0=\kappa/(1+\kappa)$ at $b=0$. To see the origin of the
equation, write $g=\gp=\gm$ at $\mu=0$. The symmetric
massless cubic and the first loop moment are
\[
 2gt_0^3+(1+\kappa-2g)t_0-\kappa=0,\qquad
 M_+=M_-=g(1-t_0^4)+\kappa(1-t_0).
\]
Eliminating $g=bM_+$ gives Eq.~\eqref{eq:neutral-massless-quintic}.
Thus the elementary quadratic for the saturation edge should
not be confused with the self-consistent neutral saddle.

Charge conjugation fixes the leading small-$\mu$ response to be odd:
\begin{equation}\label{eq:massless-density-chi}
 {
 u_*(\mu)=\chi_B\mu+O(\mu^3),\qquad
 \chi_B=
 \frac{1}{\kappa\left[
 \mathcal L_0+
 \dfrac{2b\sigma_0^2}{1-b\sigma_0(2-\sigma_0)}
 \right]}.}
\end{equation}
For the derivation, the neutral fixed-sector source curvature
in the antisymmetric direction is $-\sigma_0(2-\sigma_0)$,
the source--sector derivative is
$\sigma_0(1,-1)^{\mathsf T}$ and the sector curvature is
$-\mathcal L_0$. Differentiating the two mean-field equations
gives
\[
 \frac{\dd(\gp,\gm)}{\dd v}
 =-\frac{b\sigma_0}{1-b\sigma_0(2-\sigma_0)}
       (1,-1)^{\mathsf T}.
\]
Inserting this response into the derivative of the sector equation
gives the denominator in \eqref{eq:massless-density-chi}.
At zero coupling it reduces to
\begin{equation}
 \chi_B(0,\kappa)=
 \frac{1}{2\kappa\log[(1+\kappa)/\kappa]}.
\end{equation}
The factor $\kappa$ is fixed by $u=q/N_f$ and by the
normalization on $NN_f$. These formulas describe the response
near the origin and the position of saturation independently.
They do not replace the interior equation of state between them.

\subsection{Heavy-dense limit and the two edges}\label{sec:heavy-dense}

The heavy-dense limit keeps the quark fugacity
\[
 a=h e^\mu=e^{\mu-m}
\]
fixed while $m,\mu\to+\infty$. The antiquark fugacity
$he^{-\mu}$ vanishes. This is different from putting
$h=0$ at fixed $\mu$ in the shifted matrix integral:
the latter would discard the fugacity responsible for finite filling.

For fixed $\kappa$ and $b<1$ the critical equations simplify
considerably. Define
\[
 K_0=1+\kappa+J,\qquad
 W=\frac{K_0+\sqrt{K_0^2-4J}}2,\qquad
 \Delta_{\rm hd}=\log W-\kappa\log(1-W^{-1})+\frac{J}{W}.
\]
Taking $x=h^2\to0$ in the lower and upper polynomials gives
\[
 Jw_-^2-K_0w_-+1=0,\qquad
 w_+^2-K_0w_++J=0.
\]
Their physical roots satisfy $w_+=W$, $w_-=W^{-1}$.
Consequently,
\begin{equation}\label{eq:heavy-dense-edges}
 {
 \mu_-=m-\Delta_{\rm hd}+O(e^{-2m}),\qquad
 \mu_+=m+\Delta_{\rm hd}+O(e^{-2m}).}
\end{equation}
The partial-filling interval is symmetric about $\mu=m$
at leading heavy-dense order. At finite mass the antiquark
determinant breaks this simple symmetry. The negative-density
edges are obtained by $\mu\mapsto-\mu$.

At $b=0$ there is a simple closed equation of state throughout
this interval. Put $\widehat\mu=\mu-m$ and define
\[
 \mathcal A(y)=\tfrac12 y^2\log y,\qquad \mathcal A(0)=0.
\]
The entropy of the fixed baryon sector is
\begin{align}
 s_{\rm hd}(u)=\frac1\kappa\bigl[
 &\mathcal A(1+\kappa)-\mathcal A(1+\kappa u)
 -\mathcal A(1+\kappa(1-u))\notag\\
 &+\mathcal A(\kappa u)+\mathcal A(\kappa(1-u))
 -\mathcal A(\kappa)\bigr].
 \label{eq:hd-entropy}
\end{align}
It is symmetric under $u\mapsto1-u$ and vanishes at $u=0,1$.
At finite $N$ it follows from the hook formula for
$s_{(N^q)}(1^{N_f})$, the coefficient of $a^{Nq}$ in the
single-determinant $SU(N)$ integral. Taking $q/N_f=u$ and
the leading large-$N$ logarithm gives
Eq.~\eqref{eq:hd-entropy}. Hence
\begin{equation}
 f_*=m+\max_{0\le u\le1}
       [s_{\rm hd}(u)+\widehat\mu u]+o(1).
\end{equation}
In the intermediate phase the extremum condition reads
\begin{align}
 \widehat\mu={}&(1+\kappa u)\log(1+\kappa u)
 -(1+\kappa(1-u))\log(1+\kappa(1-u))\notag\\
 &-\kappa u\log(\kappa u)
 +\kappa(1-u)\log(\kappa(1-u)).
 \label{eq:hd-state}
\end{align}
It gives
$\Delta_{\rm hd}=(1+\kappa)\log(1+\kappa)-\kappa\log\kappa$
at $b=0$, in agreement with \eqref{eq:heavy-dense-edges}.
It also shows explicitly how a logarithm enters the
approach to an endpoint:
\[
 s_{\rm hd}''(u)=
 \kappa\log\frac{\kappa^2u(1-u)}
 {(1+\kappa u)(1+\kappa(1-u))}<0.
\]

For example, let $\delta=1-u$ and
$\Delta\mu=\mu_+-\mu\downarrow0$ on the heavy-dense
$b=0$ slice. Then
\[
 \Delta\mu=\kappa\delta\log\frac{C_{\rm hd}}{\delta}
                +O(\delta^2),\qquad
 C_{\rm hd}=\frac{e(1+\kappa)}{\kappa^2}.
\]
Inverting the leading term selects the lower real Lambert branch:
\begin{equation}\label{eq:hd-lambert}
 {1-u_*\sim
 -\frac{\Delta\mu}{
 \kappa W_{-1}[-\Delta\mu/(\kappa C_{\rm hd})]}.}
\end{equation}
The analogous onset formula follows from $u\mapsto1-u$
and $\widehat\mu\mapsto-\widehat\mu$.
The density is continuous at either edge and its
susceptibility tends to zero logarithmically. The limiting
singularity is therefore not a density jump at finite
$\kappa$; it must be distinguished from the first-order
dilute-flavor limit discussed below. Equation
\eqref{eq:hd-lambert} is stated on the $b=0$ heavy-dense
slice, where its logarithmic constant has been obtained
explicitly; changing the coupling requires the regular
part of the self-consistent sector curvature as well.

\subsection{Many flavors and free static fermions}\label{sec:many-flavors}

A different simplification occurs for $\kappa\to\infty$
at fixed finite $m,\mu$ and $b$. The determinant term then
dominates the action. The eigenvalues concentrate near the
identity; to leading order the two original Polyakov loops
approach one. Evaluating the restored determinant there gives
\begin{align}
 f_*&=\log(2\cosh m+2\cosh\mu)
          +O(\log\kappa/\kappa),\label{eq:many-flavor-free}\\
 u_*&\longrightarrow
 \frac{\sinh\mu}{\cosh m+\cosh\mu},\qquad
 \Sigma\longrightarrow
 \frac{\sinh m}{\cosh m+\cosh\mu}.
 \label{eq:many-flavor-observables}
\end{align}
The estimate for the free energy holds for fixed parameters
in a regular neighborhood of this saddle. The Vandermonde
and the spatial interaction give subleading contributions
in the $NN_f$ normalization. In particular, at $m=0$ the
limiting density is $\tanh(\mu/2)$ and its initial
susceptibility is $1/2$, also recovered from
\eqref{eq:massless-density-chi}.

There is no saturation boundary at a finite chemical
potential in this limiting expression. The finite-$\kappa$
upper edge moves to infinity. Indeed, with fixed $m$,
the upper cubic gives
\[
 w_+=\frac{\kappa(1+h^2)}{1-b}+O(1),\qquad
 \mu_+=1+\log\frac{2\kappa\cosh m}{1-b}+o(1).
\]
Thus the finite-state saturation bound remains present at
every finite $\kappa$, while at any fixed chemical potential
one sees the free-fermion profile in the many-flavor limit.
This limit is also distinct from the GWW scaling, in which
$h\to0$ while $\kappa h$ is kept finite.

\subsection{The pure-gauge limit and the threshold at \texorpdfstring{$|\mu|=m$}{absolute mu = m}}
\label{sec:quenched-mf}

In pure glue the appropriate mean-field logarithm per $N^2$ is
$\Psi_{\rm YM}(g)-g^2/b$. Inserting \eqref{eq:quenched-moments}
into \eqref{eq:mf-sources} gives $g=b\mathrm w(g)$ for $g>0$.
Thus for every fixed $0\le b<1$ the pure-glue mean-field solution
has $\gp=\gm=0$ and vanishing loops; it has no dependence on $\mu$ or
$m$. At $b=1$ the pure-glue action is flat over
$0\le g\le1/2$. The semiclassical GWW gap in the problem at
\emph{fixed sources} therefore has a different meaning from this
degenerate mean-field point, where the sources are dynamical.

The density thresholds have a well-defined small-flavour limit
for fixed $m>0$, $b<1$. With $c=b/(1-b)$ and $x=e^{-2m}$, expanding
the two physical roots gives $w_-=1-\kappa+O(\kappa^2)$ and
$w_+=1+\kappa+O(\kappa^2)$. Hence
\begin{align}
 \mu_- &= m+\kappa\log\kappa
    -\kappa[1+\log(1-x)+c(1-x)]+O(\kappa^2),\\
 \mu_+ &= m-\kappa\log\kappa
    +\kappa[1-\log(1-x)+c(1+x)]+O(\kappa^2),\\
 \mu_+-\mu_- &=2\kappa[\log(1/\kappa)+1+c]+O(\kappa^2).
 \label{eq:mf-quenched-width}
\end{align}
These expansions are not uniform in $m$ or $b$. When $m=0$ the two
determinant poles coincide; there is no lower line and
$\mu_+=2\kappa[\log(1/\kappa)+1-\log 2+c]+O(\kappa^2)$.
Let $f_*=\lim (VNN_f)^{-1}\log Z_\Lambda$ denote the
pressure on the continued saddle in the normalization used throughout
this section. For every fixed $|\mu|\ne m$ and all sufficiently small
$\kappa$, the saddle lies on one of the two plateaux. On the neutral
plateau $f_*=m+\kappa[-\log(1-x)+cx]$, whereas in saturation
$f_*=|\mu|$. Consequently,
\begin{equation}\label{eq:dilute-quark-threshold}
 {\displaystyle
 \lim_{\kappa\downarrow0}f_*=\max(m,|\mu|),\qquad
 \lim_{\kappa\downarrow0}n_B=
 \begin{cases}
  0,&|\mu|<m,\\
  \operatorname{sgn}(\mu),&|\mu|>m .
 \end{cases}}
\end{equation}
The pressure formula extends to $|\mu|=m$ by continuity:
on the intervening branch $|\partial_\mu f_*|\le1$ and
the interval between the plateaux shrinks to zero. The density
formula specifies the two one-sided limits, without assigning a
density on the threshold itself. The pressure has a kink at each
of $\mu=\pm m$ for $m>0$, and the density jumps by one in magnitude.
The same plateau formulas give $\lim_{\kappa\downarrow0}\Sigma=1$
for $|\mu|<m$ and $0$ for $|\mu|>m$. This is a first-order
filling threshold. No subtraction is needed:
at fixed $b<1$ the self-consistent pure-glue saddle has zero pressure
with normalized Haar measure. In contrast, setting $\kappa=0$ at
the start discards the determinant, and the $NN_f$ normalization
is then undefined. The fixed-source GWW limit of
Section~\ref{sec:quenched-site} isolates the pure-glue response and
does not give the dilute-flavor thermodynamic limit above.

\begin{figure}[tbp]
 \centering
 \includegraphics[width=.96\textwidth]{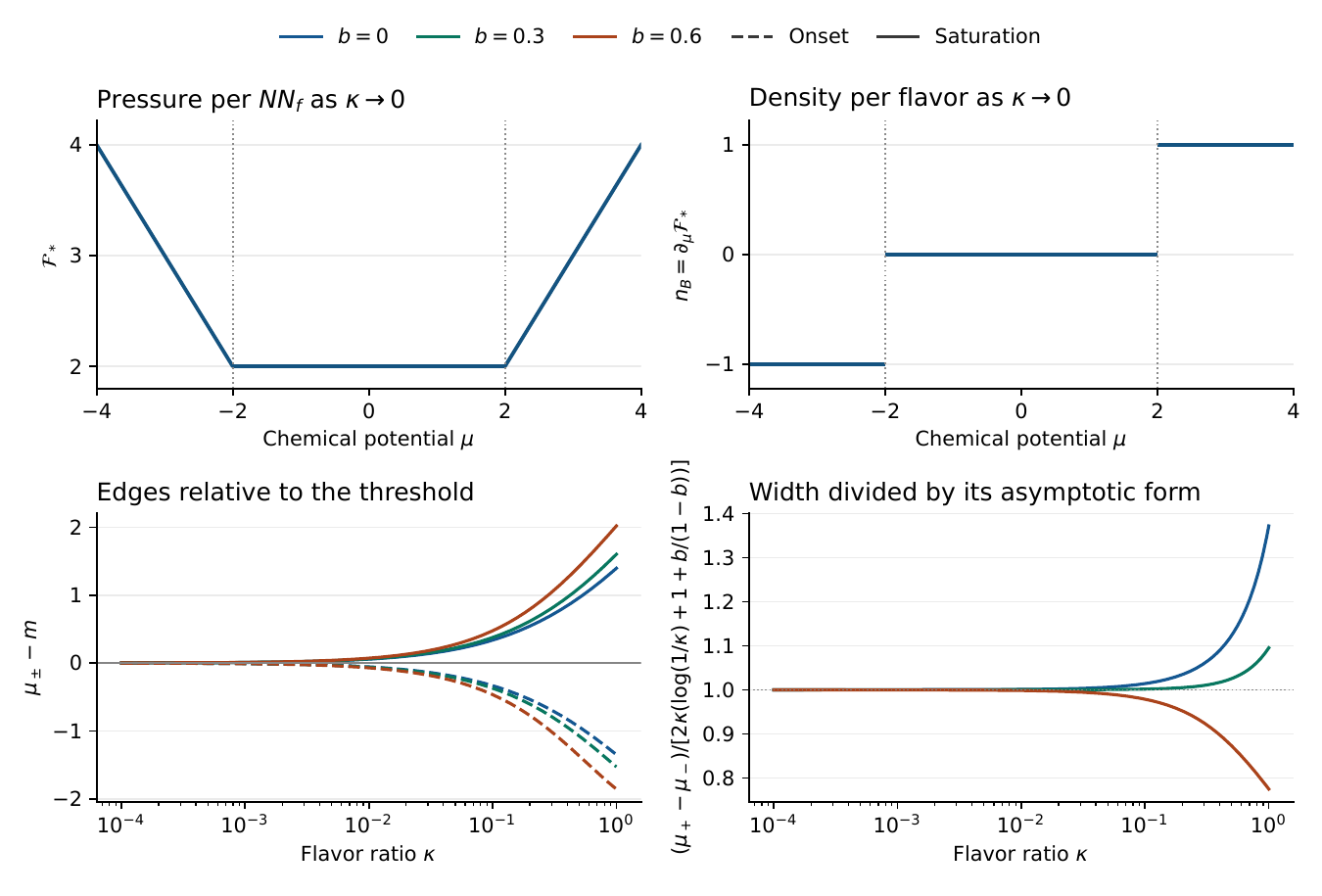}
 \caption{The dilute-flavor threshold at $m=2$. Top: the limiting
 logarithmic pressure per $NN_f$ and its $\mu$ derivative, with jumps
 at $\mu=\pm m$. Bottom left: at finite $\kappa$, the positive onset
 edge $\mu_-$ (dashed) and saturation edge $\mu_+$ (solid), both
 measured relative to $m$. Bottom right: their separation divided by
 the leading small-$\kappa$ expression in Eq.~\eqref{eq:mf-quenched-width}.}
 \label{fig:edge-collapse}
\end{figure}

Figure~\ref{fig:edge-collapse} shows the pressure and filling directly
in this limit, along with the approach of the two boundaries to
$\mu=m$ at $m=2$. Their separation agrees with
\eqref{eq:mf-quenched-width}. The limit is at fixed $b<1$.
The expansion does not apply uniformly if $b$ moves toward one
at the same time, since $J=\kappa b/(1-b)$ need not tend to zero.

\FloatBarrier
\section{Polyakov-loop screening spectrum}\label{sec:screening}

\subsection{Linearized Schwinger--Dyson equation}

Work about the homogeneous saddle of Section~\ref{sec:mf} and retain
the leading connected fluctuations, of order $N^{-2}$. Introduce
independent probing sources $j_\pm(x)$ added to
$g_\pm^{\rm phys}$ for the two original loop orientations,
with one-site exponent $N^2(j_+\mathrm w+j_-\mathrm w^\dagger)$. Let $\chi$ be the
source Hessian of the one-site $SU(N)$ logarithm in $N^2$ normalization,
at fixed chemical potential. It includes the response of the local
determinant sector, but no spatial feedback. Write
$\gamma(k)=\sum_{i=1}^d\cos k_i$ and
$\mathsf J=\left(\begin{smallmatrix}0&1\\1&0\end{smallmatrix}\right)$.
The induced field at a site is $\beta/2$ times the sum of the
opposite-orientation loops at the $2d$ neighboring sites.
The linearized Schwinger--Dyson equation and its solution are
\begin{align}
 \delta L(k)&=\chi[\delta j(k)+\beta\gamma(k)\mathsf J\delta L(k)],\\
 N^2 C(k)&=[I-\beta\gamma(k)\chi\mathsf J]^{-1}\chi .
 \label{eq:screen-dyson}
\end{align}
Here $\delta L=(\delta\mathrm w,\delta\mathrm w^\dagger)^\mathsf T$ and $C$ is
the connected spatial two-point function. This is the two-channel
susceptibility form of the Gaussian calculation in
Ref.~\cite{BCVscreening}; see also the independent-source formulation of Ref.~\cite{GWWnote}. The homogeneous mean-field saddle
provides the local vertices; factorizing at that saddle alone
would omit the connected propagator.

For $\chi=\left(\begin{smallmatrix}a&c\\c&e\end{smallmatrix}\right)$,
set $z_k=\beta\gamma(k)$. The full propagator and the eigenvalues of
$\chi\mathsf J$ are
\begin{equation}\label{eq:screen-prop}
 N^2 C(k)=
 \frac{\begin{pmatrix}a&c-z_k(c^2-ae)\\c-z_k(c^2-ae)&e\end{pmatrix}}
 {(1-z_kc)^2-z_k^2ae},\qquad
 \lambda_\pm=c\pm\sqrt{ae}.
\end{equation}
The lattice pole parameters are $\mathfrak m_\pm=1/(\beta\lambda_\pm)-d$.
The inverse exponential lengths $\mathcal M_\pm$ along a lattice axis are
\begin{equation}\label{eq:screen-mass}
 \cosh\mathcal M_\pm=1+\mathfrak m_\pm
 =1+d\left(\frac{1}{b\lambda_\pm}-1\right).
\end{equation}
This follows by setting one momentum to $i\mathcal M_\pm$ and the
others to zero. At $b=0$ there is no spatial propagation.

The momentum is transverse-zero and the exponential length is
measured along a lattice axis. The mass $\mathcal M$ is thus a
decay rate per lattice spacing, not the additive lattice
parameter $\mathfrak m$. Near a soft mode,
$\mathcal M^2\simeq 2d(1-b\lambda)/(b\lambda)$; away from it
the exact cosh relation should be used. The eigenvalues
set the pole positions, whereas the residues determine which
linear combinations of loop observables can see each pole.
This distinction is important when two eigenvalues coincide:
a repeated factor in the determinant of the inverse propagator
need not produce a second-order pole in a correlator.

\subsection{Exact local Hessian on the one-cut saddle}

Use the shifted sources $\gp,\gm$ to compute $\widehat\chi$. With
$D_\mu=\operatorname{diag}(e^{-\mu},e^\mu)$, the original-source
susceptibility is $\chi=D_\mu\widehat\chi D_\mu$. Since
$D_\mu\mathsf J D_\mu=\mathsf J$, the two matrices
$\chi\mathsf J$ and $\widehat\chi\mathsf J$ are similar.
Differentiating $\Psi_v+\mu=0$ at fixed external $\mu$ gives the
Schur complement
\begin{equation}\label{eq:screen-schur}
 \widehat\chi=\Psi_{(\gp,\gm)(\gp,\gm)}
 -\frac{\Psi_{(\gp,\gm)v}\Psi_{v(\gp,\gm)}}{\Psi_{vv}}.
\end{equation}
Holding $u$ fixed in both source derivatives would miss the second
term and gives a different spectrum.

For example, under a variation $\delta X$ of $X=(\gp,\gm)$
at fixed $\mu$ one has
$\delta v=-\Psi_{vX}\delta X/\Psi_{vv}$. Substituting this
into $\delta(\Psi_X)$ gives \eqref{eq:screen-schur}.
The rank-one correction does not describe propagation
between sites; it describes relaxation of the local charge.
Spatial propagation enters only through $\beta\gamma(k)$.
This is also why one must differentiate the local one-site
logarithm before imposing spatial self-consistency on its
second derivatives: differentiating the full mean-field
response and then summing spatial insertions would double-count
the interaction.

On the real one-cut branch in Section~\ref{sec:mf}, $p<0$ and the
conjugate endpoints are $z_\pm=(-p)e^{\pm i\theta_0}$,
$0<\theta_0<\pi$. Define
\begin{equation}\label{eq:screen-geometry}
 \sigma=\sin^2(\theta_0/2)=\frac{1+L/p}{2},
 \qquad \mathcal L=-\log(1-\sigma).
\end{equation}
The fixed-sector Hessian and charge curvature are computed by
varying the loop equation; Appendix~\ref{app:hessian} gives
the resolvent variations and the logarithmic integral.
Substitution into \eqref{eq:screen-schur} gives
\begin{align}
 \widehat\chi &=
 \begin{pmatrix}p^2 Q&S\\S&p^{-2}Q\end{pmatrix},\qquad
 Q=\frac{\sigma^2}{\mathcal L}-\sigma(1-\sigma),
 \quad S=\sigma-\frac{\sigma^2}{\mathcal L},\\
 {\lambda_+=\sigma^2,\qquad
 \lambda_-=\sigma(2-\sigma)-\frac{2\sigma^2}{\mathcal L}.}
 \label{eq:screen-eigs}
\end{align}
The loop asymmetry enters the residues, while both masses only
depend on $\sigma$. In the neutral gapped GWW limit,
$\sigma=1/(2g)$, and \eqref{eq:screen-eigs} reduces to the
gapped $SU(N)$ response with logarithm $\log[2g/(2g-1)]$.

The fact that $p$ drops out of the eigenvalues is a consequence
of the two orientations of the interaction. Multiplication by
$\mathsf J$ exchanges the two columns of $\widehat\chi$, so
its diagonal entries become equal to $S$ and its off-diagonal
entries are $p^2Q$ and $p^{-2}Q$. A diagonal similarity
transformation reduces this to a symmetric matrix with
off-diagonal entry $Q$. Thus one combination has eigenvalue
$S+Q$ and the other $S-Q$. Expressed back in the original
basis their amplitudes will, in general, not be related by
an equal-weight sum and difference. The pole locations can
be compared without matching those amplitudes, provided both
channels are resolved and the comparison uses the same
transverse momentum and lattice mass convention.

\subsection{An explicit neutral massless slice}

At $m=\mu=0$, the solution of
\eqref{eq:neutral-massless-quintic} gives $\sigma_0=1-t_0^2$.
The two axial screening masses can therefore be evaluated from
one real equation:
\begin{equation}\label{eq:massless-neutral-masses}
 \mathcal M_\pm(0,0)=
 \operatorname{arcosh}\left[
 1+d\left(\frac1{b\lambda_\pm(\sigma_0)}-1\right)\right],
 \qquad 0<b<1.
\end{equation}
The neutral saddle itself has the simple zero-coupling values
$t_0=\kappa/(1+\kappa)$ and
$\sigma_0=(1+2\kappa)/(1+\kappa)^2$.
For small positive $b$, denote the eigenvalues at this last
$\sigma_0$ by $\lambda_\pm^{(0)}$. Then
\[
 \mathcal M_\pm=
 \log\frac{2d}{b\lambda_\pm^{(0)}}+O(b).
\]
Both masses diverge as the spatial interaction is switched off,
as expected for independent sites. Their finite separation at
small positive $b$ is fixed by the two local response channels.
At finite $b$ the source equation must be solved together with
the cut equation; replacing $t_0$ by its zero-coupling value
would miss that feedback.

\subsection{Reality of the poles and the density thresholds}

The discriminant is $4ae=4Q^2$ on this branch. In fact
\begin{equation}
 \frac{2\sigma}{2-\sigma}<-\log(1-\sigma)
 <\frac{\sigma}{1-\sigma}\qquad(0<\sigma<1).
\end{equation}
Both bounds follow by differentiation from equality at $\sigma=0$.
They give $Q>0$ and $0<\lambda_-<\lambda_+<1$. Thus both masses
are real, positive for $0<b<1$, and distinct on the regular
one-cut saddle. This excludes a complex-conjugate pair throughout
that branch, independently of the mass, sources and flavour
ratio used to reach it.

On either open density plateau the sector is pinned. The
full-support local logarithm has an $\gp\gm$ term and terms at
most linear in the sources, from the Laurent coefficients on
the closed contour. At saturation one first factors out the
largest determinant power. Thus $\widehat\chi=\mathsf J$ on
either plateau, and the two masses coincide at
$\operatorname{arcosh}[1+d(b^{-1}-1)]$. The one-cut formula
tends to this value as $\sigma\to1$. The double denominator
zero at $\lambda_\pm=1$ leaves a simple pole,
$N^2 C(k)=\mathsf J/(1-\beta\gamma(k))$. Neither screening mass
vanishes at these thresholds for $b<1$.

The approach to the common plateau mass gives another check
on the interpretation of the roots. Let $\epsilon=1-\sigma$.
Equation~\eqref{eq:screen-eigs} gives
\[
 \lambda_+=(1-\epsilon)^2,\qquad
 \lambda_-=1-\epsilon^2-
                 \frac{2(1-\epsilon)^2}{\log(1/\epsilon)}.
\]
Both tend to one, but the approach of the heavier mode is
logarithmically slow in the cut parameter. If
$\mathcal M_0=\operatorname{arcosh}[1+d(b^{-1}-1)]>0$,
their splitting is asymptotically
$2d/[b\,\sinh(\mathcal M_0)\log(1/\epsilon)]$ at fixed $b$.
This slowing concerns the opening of the cut; it does not
imply a massless spatial mode. It also makes a numerical
comparison very close to the threshold more sensitive to the
endpoint geometry than a comparison deeper inside a plateau.

\begin{figure}[htbp]
 \centering
 \includegraphics[width=\textwidth]{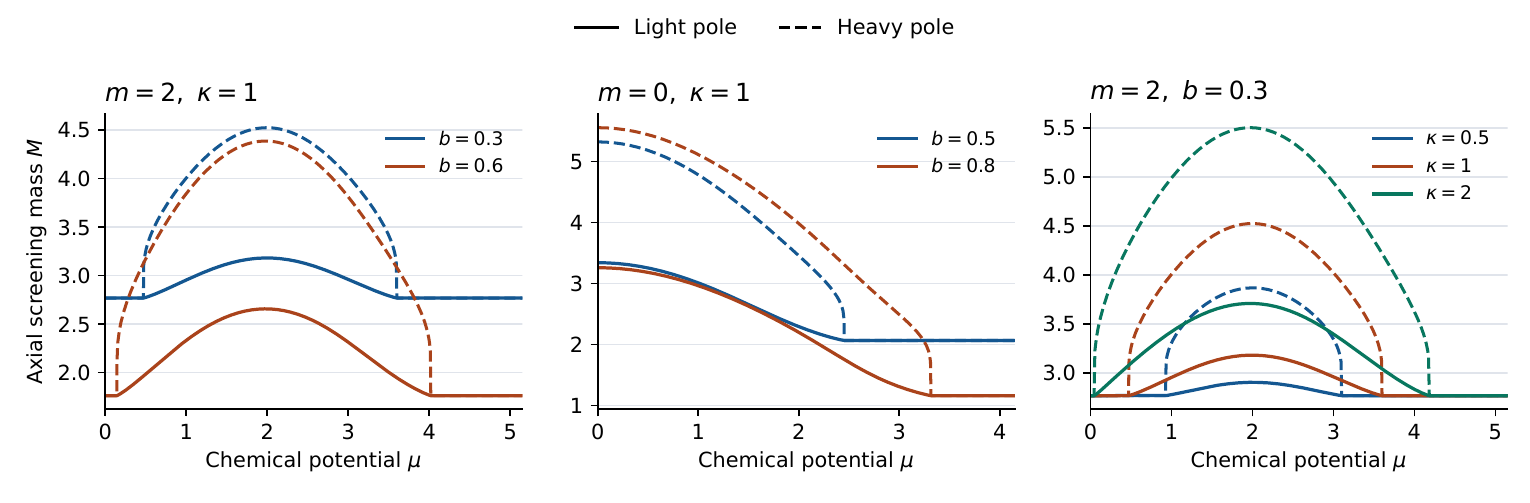}
 \caption{Axial screening masses $\mathcal M_\pm$ in lattice
units for three spatial dimensions. Left: $m=2$, $\kappa=1$
at two couplings. Center: massless cuts at $\kappa=1$.
Right: flavor comparison at $m=2$, $b=0.3$.
The lighter pole is solid and the heavier pole dashed; colors
compare parameters within each panel. The two curves coincide
on the zero-density and saturation plateaux wherever those
intervals exist. Their splitting diagnoses two distinct decay
rates on the interior branch, not the onset of an oscillatory
correlator.}
\label{fig:screening}
\end{figure}

Figure~\ref{fig:screening} shows the $d=3$ masses on six
thermodynamic cuts. For the massive $\kappa=1$ examples, the
splitting is largest near $\mu=m$; for the massless examples
it is largest at $\mu=0$. Small independent variations of
each source and of $v$ verify \eqref{eq:screen-fixed-hess};
direct quadrature verifies \eqref{eq:screen-sector-curvature}.
The result applies to the plateau and regular one-cut solutions
used in Section~\ref{sec:mf}.

For purely imaginary $\mu$, the physical sources satisfy
$j_-=j_+^*$ and the static determinants are conjugate. The
exact local measure is positive. Covariance positivity
gives $e=a^*$ and $c\ge |a|$, so
$\lambda_\pm=c\pm|a|\ge0$. On a stable branch the finite
screening rates are real; this result also holds on each
side of a switch between center sectors. For general complex
$\mu$, the real-endpoint and positivity arguments need not
apply. The same Dyson equation can be continued branch by
branch; its eigenvalues can become complex without forming
a conjugate pair at the same $\mu$. This is distinct from
an oscillatory regime of the real-density phase diagram.
For a small imaginary displacement $\eta$ about a regular
real value $\mu_0$, analyticity and reality on the real axis give
\begin{equation}\label{eq:complex-mass-linear}
 {\operatorname{Im}\mathcal M_\pm(\mu_0+i\eta)=
 -\eta\,\frac{d}{
 b\lambda_\pm^2\sinh\mathcal M_\pm}
 \frac{\dd\lambda_\pm}{\dd\mu}(\mu_0)+O(\eta^3).}
\end{equation}
The real part changes first at order $\eta^2$. All quantities on
the right are evaluated on the real self-consistent solution.
Since the two eigenvalues in \eqref{eq:screen-eigs} depend only
on $\sigma$, their derivatives follow from
$\dd\lambda_\pm/\dd\mu=(\partial_\sigma\lambda_\pm)
(\dd\sigma/\dd v)/(\dd\mu/\dd v)$. Thus an imaginary part of a
screening mass near the real axis measures the slope of its
real-density profile, rather than a new independent parameter.

For instance, continuing the saddle from $\mu=2.5$ to
$\mu=2.5+0.1i$ at $(m,\kappa,b,d)=(2,1,0.3,3)$ gives
\[
 \mathcal M_+\simeq 3.12036-0.02426i,\qquad
 \mathcal M_-\simeq 4.39549-0.05202i.
\]
The continuation uses $\dd\mu/\dd v=\mathcal L-a_v^\mathsf T
\dd(\gp,\gm)^\mathsf T/\dd v$, where $a_v=\Psi_{(\gp,\gm)v}$ and the
source tangent follows from \eqref{eq:mf-sources} at fixed $b$.
Continuing to $2.5-0.1i$ conjugates each pole individually.
The same continuation about the conjugate point gives the
complex-conjugate response.

\FloatBarrier
\section{Imaginary chemical potential and the \texorpdfstring{$SU(3)$}{SU(3)} benchmark}
\label{sec:rw}

Put $\mu=i\theta$ and define the center angle $\phi=N\theta$.
A global center rotation compensates
$\theta\mapsto\theta+2\pi/N$, and charge conjugation takes
$\theta\mapsto-\theta$. Thus the appropriate window for one
Roberge--Weiss period~\cite{RW} is $0\le\phi\le2\pi$.
The boundary $\phi=\pi$ is invariant under the combination of
these two symmetries. Periodicity fixes this candidate boundary;
it does not imply a transition there.

In the shifted sources write $\gp=x+iy$ and $\gm=x-iy$.
Let $c_j$ be the Fourier coefficients of the real nonnegative
weight $w(\omega)$:
\[
 w(\omega)=e^{2N(x\cos\omega-y\sin\omega)}
       (2\cosh m+2\cos\omega)^{N_f},\qquad
 c_j=\int_{-\pi}^{\pi}\frac{\dd\omega}{2\pi}\,
       e^{-ij\omega}w(\omega).
\]
The exact finite-$N$ local integral and homogeneous pressure are
\begin{align}
 f_N(x,y;\phi)&=\frac{1}{NN_f}\log
    \sum_{q\in\mathbb Z}e^{iq\phi}
       \det[c_{q+i-j}]_{i,j=1}^N,\label{eq:rw-toeplitz}\\
 P_N(\phi)&=\max_{x,y}
      \left[f_N(x,y;\phi)-\frac{x^2+y^2}{\kappa b}\right] .
 \label{eq:rw-mf}
\end{align}
For $b=0$ the second line means $x=y=0$. The counterterm has
the same $NN_f$ normalization as in \eqref{eq:mf-action}.
In the original matrix variable the
local measure is positive. In finite volume the partition
function is smooth in $\phi$; a cusp in \eqref{eq:rw-mf} arises
only when choosing the dominant homogeneous branch. This
construction tests mean field, and does not prove a transition
of the unrestricted spatial model at these finite $N$ values.

In \eqref{eq:rw-toeplitz}, the shift of Fourier indices implements
the same determinant constraint as in the real-density
calculation. The angle dependence, however, is now in a phase
$e^{iq\phi}$ rather than a positive fugacity factor. At
$\phi=\pi$, conjugation followed by a center transformation sends
$y$ to $-y$. A single smooth extremum fixed by this operation
has zero angle derivative there. A pair of dominant extrema
related by this operation can instead have opposite one-sided
derivatives. Maximizing only along the $y=0$ axis would miss the
second possibility, even though it would preserve the location
and periodicity of the center boundary.

\paragraph{What survives at large $N$.}
On a regular gapped neutral saddle write $x=x_0$ at leading
order and $y=t/N$. The modes that resolve a single period
have $q=O(1)$, or $v=q/N=O(N^{-1})$. The planar Gaussian
expansion of the fixed-sector action gives their relative
weights $\exp[-\mathcal L q^2/2+2i\sigma t q]$:
$\Psi_{vv}=-\mathcal L$ and $\Psi_{yv}=2i\sigma$ at $p=-1$.
The fixed-sector $y$ curvature is $2\sigma(2-\sigma)$.
Consequently, in this regular expansion the center dependence
of $NN_fP_N$ is given, up to an angle-independent constant, by
the maximum over $t$ of
\begin{equation}\label{eq:rw-scaled}
 -\bigl[b^{-1}-\sigma(2-\sigma)\bigr]t^2+
 \log\Theta_{\mathcal L}(\phi+2\sigma t),\qquad
 \Theta_{\mathcal L}(s)=\sum_{q\in\mathbb Z}
           e^{-\mathcal L q^2/2+iqs}.
\end{equation}
Here $\sigma$ is evaluated at the neutral planar saddle and
$\mathcal L=-\log(1-\sigma)$. This approximation is for a regular
cut, away from changes of its topology. In the full-support region
with positive charge threshold, nonzero fixed $q$ are instead
exponentially suppressed in $N$. In particular, a cusp of
\eqref{eq:rw-scaled} has an order-one slope discontinuity in
the scaled pressure as a function of the center angle. The variation
of $P_N$ itself is $O((NN_f)^{-1})$ in this window, and the
normalized density $\partial_\mu P_N$ is $O(N_f^{-1})$, since
$\phi=N\,\operatorname{Im}\mu$. Replacing the integer
sum by a continuous saddle before taking this limit loses its
periodicity.

The two $O(1)$ effects in \eqref{eq:rw-scaled} come from
different sources. The angular displacement $t/N$ costs a
quadratic term in the homogeneous action, while the fixed-charge
curvature weights each \emph{integer} $q$ separately. The sum
has the equivalent form
\begin{equation}
 \Theta_{\mathcal L}(s)
 = \sqrt{2\pi/\mathcal L}\,
    \sum_{j\in\mathbb Z}\exp[-(s-2\pi j)^2/(2\mathcal L)] .
\end{equation}
This makes positivity and periodicity explicit. It also shows
why extending the integral over a continuous charge and keeping
only a single Gaussian is inadequate over an entire period.
The expression holds as a regular one-cut expansion; it is not
a replacement for the exact finite-$N$ Toeplitz sum when the
cut approaches a degeneration.

\paragraph{Screening at and across the center boundary.}
Evaluate the real $(x,y)$ Hessian $H_N=\nabla^2 f_N$ at the
selected saddle, before adding the counterterm or differentiating
the saddle with respect to the source. The eigenvalues of
$\chi\mathsf J$ in \eqref{eq:screen-prop} are the two eigenvalues
of $H_N/2$. Thus they are nonnegative, and $b\lambda_\pm<1$ on a
strictly stable saddle. The axial masses are real and follow
from \eqref{eq:screen-mass}. At a cusp the two saddles are related
by $y\mapsto-y$: their pressure slopes in $\phi$ are opposite but
their Hessians have identical eigenvalues. The two \emph{ordered
masses} have equal left and right limits, even though their slopes
can differ and the two loop orientations are interchanged.

Equation~\eqref{eq:rw-scaled} also displays the two local responses
in the scaling window. With $s=\phi+2\sigma t_*$ on the selected
branch, the real Hessian becomes diagonal at leading order and
the radial and angular eigenvalues are
\begin{equation}\label{eq:rw-lambda}
 \lambda_{\mathrm{rad}}=\sigma^2,\qquad
 \lambda_{\mathrm{ang}}=\sigma(2-\sigma)+
     2\sigma^2(\log\Theta_{\mathcal L})''(s).
\end{equation}
The angular channel softens at a continuous loss of stability
of the center-symmetric saddle when $b\lambda_{\mathrm{ang}}=1$.
For example, at $m=0,\kappa=1$ and $\phi=\pi$, the condition
evaluated at $t=0$ gives $b\simeq0.15440$ in
\eqref{eq:rw-scaled}; the exact local finite-$N$ value at $N=6$
is $0.15715$ (and $0.16762$ at $N=3$). These values locate the \emph{local instability}
of that saddle, without requiring a global phase classification.

\begin{figure}[htbp]
\centering
\includegraphics[width=\textwidth]{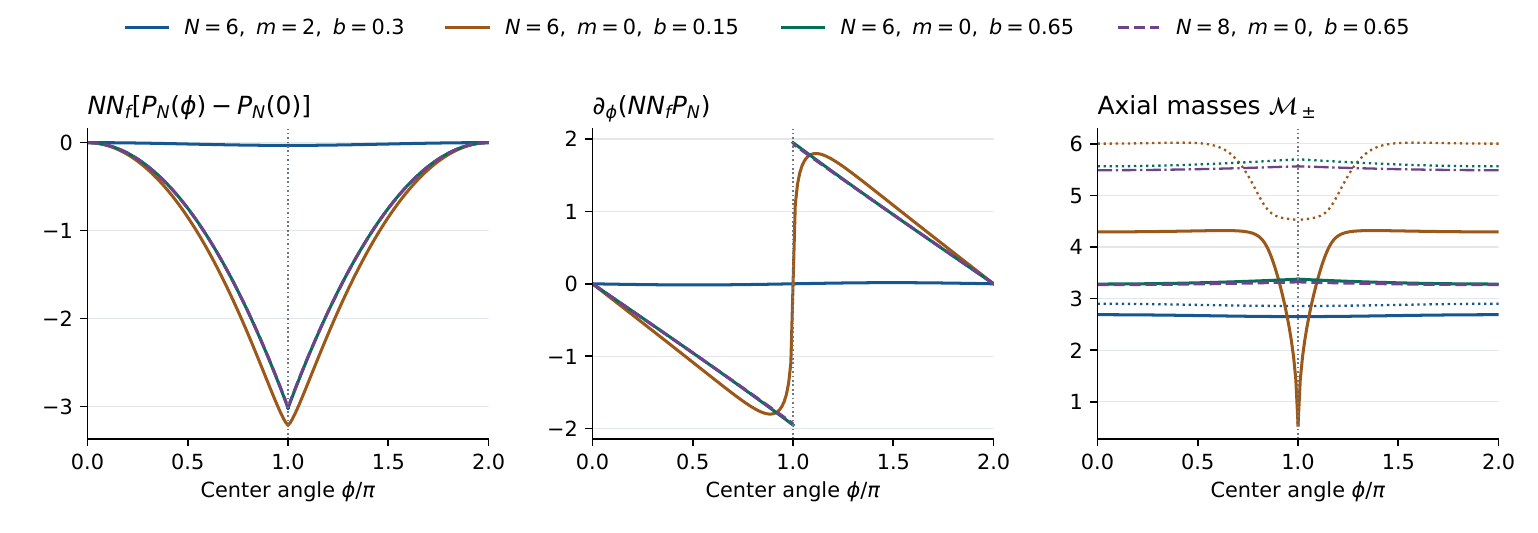}
\caption{One center period $0\le\phi\le2\pi$ at $\kappa=1$.
Left: $NN_f[P_N(\phi)-P_N(0)]$; center: its derivative with respect
to $\phi$, equal to $i\langle B_{\rm tot}\rangle/V$ in the pressure
convention used here. The jump at $\phi/\pi=1$ occurs on the
massless $b=0.65$ examples, not on the weaker or massive cuts
shown. Right: the two axial masses in $d=3$ in lattice units;
the finer dash pattern indicates the heavier mass in each pair.
The ordered masses agree on the two sides of the boundary
despite the jump in the angle derivative. All curves use
the exact finite-$N$ one-site determinant in homogeneous mean field.}
\label{fig:rw}
\end{figure}

For the two massless cuts at $b=0.65$ in Fig.~\ref{fig:rw},
the pressure has a cusp and its scaled slope jumps by about $3.90$
at $N=6$ and $3.86$ at $N=8$; the regular expansion
\eqref{eq:rw-scaled} gives $3.82$ for this jump. The one-sided masses are,
respectively, $(3.376,5.700)$ and $(3.323,5.564)$ in lattice units.
By contrast, the exact $N=6$ massless cut at $b=0.15$ is smooth
and its lighter mass falls to $0.549$ at the boundary; the
massive example is smooth as well. The local determinant quadratures
were repeated with a larger Fourier and sector cutoff. These examples
exhibit both the surviving cusp at fixed $\phi$ and a real spectrum
throughout the imaginary-potential cuts.

The radial and angular responses in \eqref{eq:rw-lambda}
also clarify what the stability criterion measures. The
local Hessian acts before the spatial feedback is included;
in the uniform angular channel the curvature of the
homogeneous variational functional is proportional to
$\lambda_{\rm ang}-1/b$. The limit $\mathcal M\to0$ is
therefore a statement about that particular stationary
solution. On a finite spatial lattice the exact pressure at
the reflection-symmetric angle has zero signed derivative.
The opposite derivatives in a cusp are obtained by
approaching the boundary within each selected phase or by
measuring the two orientations separately. Averaging the
signed derivatives of both orientations tests the symmetry
rather than the jump.

\subsection{Comparison with the spatial \texorpdfstring{$SU(3)$}{SU(3)} model}

It is helpful to separate the finite-color effect from the
homogeneous mean-field approximation. We first solve
\eqref{eq:rw-mf} at $N=3$ with the exact one-site determinant,
and compare with \eqref{eq:rw-scaled} at the corresponding
$\kappa=N_f/3$. We then compare both with direct simulations of
\eqref{eq:spatial-action} at $N=N_f=3$. No rescaling of the
coupling is applied between these three calculations.
Let $\mathcal P=NN_f P=(\log Z_\Lambda)/V$, with $V$ the number
of spatial sites. It is the reduced pressure per site, or
$-F/(TV)$, rather than a four-dimensional pressure in physical
units. The derivative along the center angle is
$\partial_\phi\mathcal P=i\langle B_{\rm tot}\rangle/V$, where
$B_{\rm tot}$ is total baryon number, and in a homogeneous
calculation $\mathcal P=NN_fP_N$. The additive normalization
cancels in the difference between angles.

\begin{table}[tbp]
\centering
\small
\setlength{\tabcolsep}{6pt}
\begin{tabular}{@{}c c rr rr@{}}
\toprule
 & & \multicolumn{2}{c}{$\mathcal P(\pi)-\mathcal P(0)$}
 & \multicolumn{2}{c}{slope jump at $\phi=\pi$}\\
$N_f$ at $N=3$ & $\kappa$ & $SU(3)$ MF & large $N$ & $SU(3)$ MF & large $N$\\
\midrule
1 & $1/3$ & $-1.178$ & $-1.214$ & $1.611$ & $1.544$\\
2 & $2/3$ & $-2.147$ & $-2.112$ & $2.920$ & $2.689$\\
3 & $1$   & $-3.085$ & $-3.000$ & $4.178$ & $3.820$\\
\bottomrule
\end{tabular}
\caption{Finite-$N$ homogeneous mean field (MF) versus the
scaled large-$N$ theory for the same static parameters $m=0$,
$b=0.65$. The slope jump is the absolute difference between
the two one-sided values of $\partial_\phi\mathcal P$.}
\label{tab:rw-thermo}
\end{table}

The reduced pressure variation in Table~\ref{tab:rw-thermo}
changes by at most $3.2\%$ between the two homogeneous
calculations, and the slope jump by at most $9\%$. In the
spatial $N_f=3$ runs on $6^3$ and $10^3$, the slope magnitude
at $\phi=0.8\pi$ is approximately $1.5967$, versus $1.5965$
at $SU(3)$ mean field and $1.5279$ at large $N$. Separate
initializations in the two sectors at $\phi=\pi$ give slopes
approximately $\pm2.089$ in the spatial model. They are
phase-resolved values, not a nonzero derivative of the
finite-volume, symmetry-restored pressure. We use these runs
to test the pressure slope; the pressure differences in
Table~\ref{tab:rw-thermo} come from the homogeneous calculations.

For the screening calculation we use the connected wall
correlator of the two components of
$\mathrm w_\theta=e^{i\theta}\Tr U/3$. We project the real symmetric
$2\times2$ correlator with the generalized eigenvectors of
$C(1)$ relative to $C(0)$, and use the periodic cosh ratio on
successive separations. The two eigenvectors provide projections onto the loop
combinations that are best resolved at the first two
separations. For a single exponential on a periodic spatial
lattice of extent $L$, the projected ratio is
\begin{equation}\label{eq:rw-cosh-ratio}
 \frac{C_v(r+1)}{C_v(r)}=
 \frac{e^{-\mathcal M(r+1)}+e^{-\mathcal M(L-r-1)}}
      {e^{-\mathcal Mr}+e^{-\mathcal M(L-r)}} ,
 \qquad C_v(r)=v^\mathsf T C(r)v.
\end{equation}
We invert this relation for each projected channel. If there
are more contributing length scales, two successive ratios
need not give the same $\mathcal M$. We therefore quote both
the distance-window comparison and its statistical error,
and describe the measured quantities as effective masses.
The mean-field masses in the same table, in contrast, are
the poles of the homogeneous spatial response.

\begin{figure}[tbp]
\centering
\includegraphics[width=.96\textwidth]{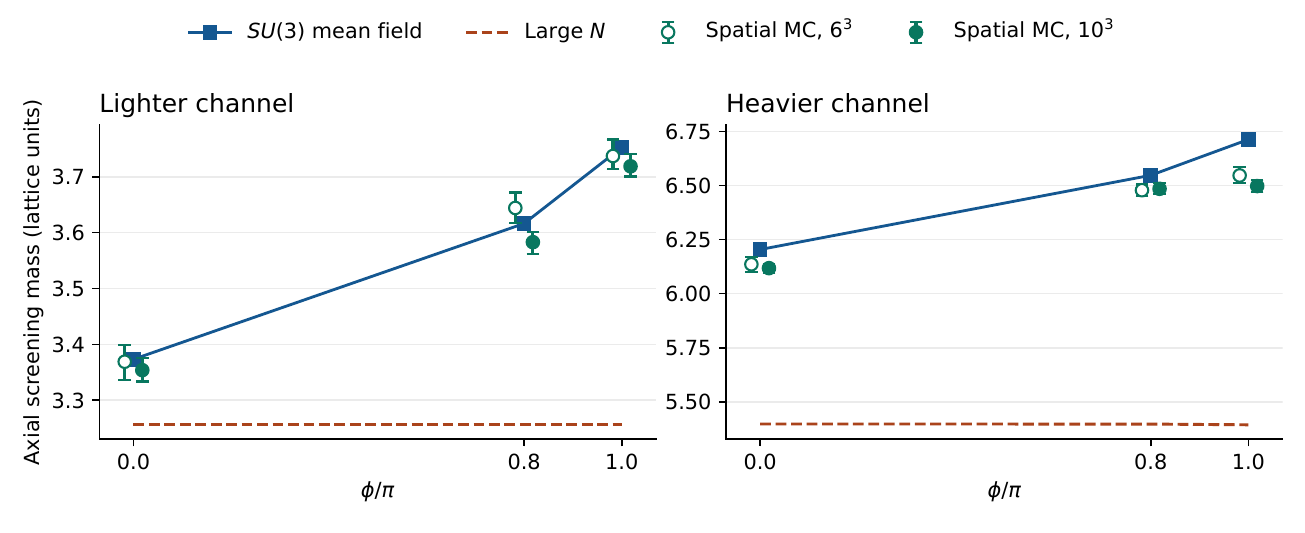}
\caption{Spatial $SU(3)$ effective masses at $m=0$,
$b=0.65$, $N_f=3$, from separations $r=1,2$, compared with
homogeneous $SU(3)$ and the scaled $\kappa=1$ result.
Left and right show the lighter and heavier channels.
The open and filled circles denote $6^3$ and $10^3$;
Monte Carlo points at the same angle are offset horizontally
for visibility. Error bars are statistical. The point at
$\phi/\pi=1$ is measured in the phase reached from below,
and lines simply connect the three sampled angles.}
\label{fig:rw-compare}
\end{figure}

The estimates are stable under taking
a fresh basis from $C(2)$ relative to $C(1)$.
Appendix~\ref{app:MC} gives the spatial estimator,
conditional averaging and simulation parameters.
Table~\ref{tab:rw-masses} and Fig.~\ref{fig:rw-compare}
show that the finite-$N$ homogeneous spectrum is closer to
the spatial results than the scaled large-$N$ spectrum.
In particular, the lighter spatial mass grows across the
center period on the displayed branch, while the lighter
large-$N$ mode is constant on this cut.
At the RW boundary the heavier effective mass moves from
$6.600(6)$ to $6.498(27)$ when the $r=0$ point is discarded
from the ratio on $10^3$. There is no resolved asymptotic
plateau for this channel at $r=3$. The discrepancy with large
$N$ is substantially larger than this change of window.

\begin{table}[tbp]
\centering
\small
\setlength{\tabcolsep}{5pt}
\begin{tabular}{@{}c cccc@{}}
\toprule
$\phi/\pi$ & MC, $6^3$ & MC, $10^3$ & $SU(3)$ MF & large $N$\\
\midrule
$0$   & $(3.37(3),\,6.14(4))$ & $(3.354(22),\,6.118(22))$ &
$(3.373,\,6.204)$ & $(3.256,\,5.397)$\\
$0.8$ & $(3.64(3),\,6.48(3))$ & $(3.583(21),\,6.485(25))$ &
$(3.616,\,6.547)$ & $(3.256,\,5.397)$\\
$1^-$ & $(3.74(3),\,6.55(4))$ & $(3.719(20),\,6.498(27))$ &
$(3.752,\,6.713)$ & $(3.256,\,5.393)$\\
\bottomrule
\end{tabular}
\caption{Lighter and heavier axial screening masses in spatial lattice
units at $m=0,\ b=0.65,\ N_f=3$ for $SU(3)$ and $\kappa=1$
for large $N$. Monte Carlo entries are effective masses from
$r=1,2$; errors in parentheses are statistical and do not
include the choice of distance window.}
\label{tab:rw-masses}
\end{table}

The comparison distinguishes three operations. First, at the
same $N=3$, chemical-potential angle and bare coupling, replacing
the spatial model by homogeneous mean field changes spatial
fluctuations while keeping the local static determinant exact.
Second, changing from $N=3$ to the scaled large-$N$ saddle at
fixed $\kappa$ changes the local response and the set of
center-sector weights, even within homogeneous mean field.
Third, replacing the static action by an action with spatial
quark propagation changes the underlying theory. The first
two comparisons can be carried out with the data here. They
do not isolate the third effect.

For example, at $\phi=0.8\pi$ the heavy effective mass on
$10^3$ is $6.485(25)$; finite-$N$ mean field gives $6.547$
and scaled large $N$ gives $5.397$. The corresponding light
masses are $3.583(21)$, $3.616$ and $3.256$. This comparison is
not normalized to one of the masses, and no coupling was
adjusted to remove a discrepancy. On moving from $\phi=0$ to
the center boundary, the light spatial mass changes by about
$0.37$, compared with about $0.38$ in finite-$N$ mean field.
On this scaled branch the light large-$N$ pole is instead
the radial mode in \eqref{eq:rw-lambda}, constant at leading
order in the center angle. These numerical contrasts thus
test both an absolute scale and a change with angle. They
also show why agreement in the pressure variation does not
imply agreement of its source Hessian.

The nearby RW endpoint would require a different comparison.
A vanishing homogeneous pole is determined by a local stability
condition, whereas a spatial critical length is extracted from
the volume and distance dependence of connected correlators.
The two-volume measurements in Table~\ref{tab:rw-masses} were
performed inside the regime in which the two orientations can
be resolved, not as a finite-size scaling study of that endpoint.

\subsection{Relation to lattice QCD}

The RW scaling can also be checked analytically in a distinct
regime. Expanding the one-loop massless-quark potential of
Ref.~\cite{Kashiwa} around center-related perturbative vacua gives
$\Delta(p/T^4)\to-(\kappa/6)\min_{j\in\mathbb Z}(\phi-2\pi j)^2$ at
fixed $\phi=N\theta$ and fixed $\kappa$. This has the same
$N^0$ scaling in the unnormalized pressure; the coefficient
is not a matching prediction for the static spin action.
An earlier GWW-based susceptibility calculation also examines
screening over a center period \cite{GWWnote}. The published
dual $SU(3)$ studies of Refs.~\cite{DualPhase,DualMass} use the
same exponential nearest-neighbour interaction and static
one-flavor determinant, at real chemical potential. They
provide a direct precedent for the spatial correlator method.
The heavy-quark theory of Ref.~\cite{Fromm} reaches imaginary
chemical potential with a different nearest-neighbour weight
and the static Wilson power $2N_f$. Both differences matter
for a quantitative match.

For full lattice QCD, Ref.~\cite{deFP} finds light- and heavy-mass
tricritical points with three degenerate dynamical flavors.
Ref.~\cite{Andreoli} measures screening at imaginary chemical
potential from a matrix of Polyakov-loop correlators for
physical $2+1$ flavors. These are qualitative comparisons:
our $m=0$ is the massless value of a static parameter, and
the action lacks spatial quark hopping. In particular, the
loss of local angular stability in \eqref{eq:rw-lambda} does
not by itself locate the spatial RW endpoint.

\section*{Summary and outlook}\label{sec:summary}
\addcontentsline{toc}{section}{Summary and outlook}

We have studied the $SU(N)$ Polyakov loop model with the exact
static quark determinant in the 't Hooft--Veneziano limit.
The local matrix integral was solved with two independent sources,
and its moments were used to close the spatial mean-field equations.
This preserves the distinction between the two loop orientations
and determines the baryon sector together with the sources.

The real-density phase diagram extends the three-phase structure
of one-dimensional QCD to nonzero spatial coupling. A lower
boundary separates the zero-density phase from partial filling;
an upper boundary marks saturation. Their exact equations are
a reciprocal quartic and a cubic, respectively. The lower
quartic reduces to a quadratic in $w+h^2/w$. When a positive
lower boundary exists, the two chemical potentials satisfy
$0<\mu_-<m<\mu_+$. Increasing $b$ lowers the onset boundary
and raises the saturation boundary. The lower boundary ends
at $\mu=0$ rather than intersecting the upper one.

The explicit limits provide useful checks and physical
interpretations. We obtained a closed massless saturation
formula and the small-$\mu$ susceptibility, including the
response of the mean field. In the heavy-dense regime the
two edges are symmetric about $\mu=m$ at leading order;
at zero coupling the complete equation of state and its
Lambert-$W_{-1}$ endpoint behavior are explicit.
As $\kappa\to0$ at fixed $b<1$, the free energy per $NN_f$
approaches $\max(m,|\mu|)$, with first-order filling thresholds
at $\mu=\pm m$. As $\kappa\to\infty$ at fixed parameters it
instead tends to the free static-fermion expression
$\log(2\cosh m+2\cosh\mu)$.

The free energy, density, two Polyakov loops and chiral
condensate were calculated on common cuts. In the zero-density
phase the free energy and mass derivative are independent of
$\mu$, although the two loop orientations have opposite
fugacity factors. Saturation fixes the filling at its maximum,
the free energy at $|\mu|$ and the condensate at zero. The
intermediate phase is resolved directly by the sector saddle.

Linearized Schwinger--Dyson equations give the connected
spatial correlator in terms of the local source Hessian.
The Hessian must include relaxation of the determinant sector.
The resulting two masses depend on the cut opening, are real
on the regular real-density saddle, and coincide on the
density plateaux. Their approach to the common plateau mass
is logarithmically slow. A small continuation into complex
chemical potential gives an imaginary part proportional to
the derivative of the real-axis mass profile.

At imaginary chemical potential the shrinking center period
is resolved by keeping $\phi=N\,\mathrm{Im}\,\mu$ fixed.
Retaining the integer sector sum gives the theta-function
description of the Roberge--Weiss signal. At the same static
action and bare couplings, the large-$N$ pressure variation
is closer to the $SU(3)$ measurements than the screening
spectrum is. Finite-$N$ mean field gives a better description
of the measured short-distance screening masses. Longer
distance correlators and a finite-size study of the spatial
endpoint would sharpen this comparison. Including spatial
quark propagation is necessary before making a quantitative
comparison to dynamical QCD.

The general algebraic response and the source-dependent
projective transformations support these calculations but
are not required to use the critical-line formulas. The
mass derivative has generic degree 38. Within the regular
lifts described in Appendix~\ref{sec:fullsym}, the source-free
$S_4$ reduces to its order-eight subgroup $D_4$.
The detailed elimination and the special lower-degree slices
are collected in the appendices.

\clearpage
\appendix
\section{Algebraic elimination and special slices}\label{app:elimination}
For compact algebraic formulas in the appendices we write
$A=g_+$ and $B=g_-$; these are the shifted sources of the main text.
Throughout this appendix $d$ is shorthand for the pole parameter
$d_{\mathrm p}=1+\kappa-v$ of Section~\ref{sec:fixed-sector},
while $e=1+\kappa+v$.

For generic nonzero sources and $0<h<1$, eliminating the signed
endpoint data gives a degree-38 equation for the auxiliary parameter
$p$. The mass response is rational in $p$ once the other endpoint
data are reconstructed, and its monic equation has the same generic
degree. Appendix~\ref{app:elimination} gives the two cubics entering
the resultant, the factor removed at $p=0$ and a finite-field
certificate that the degree is attained. The polynomial encodes all
the regular one-cut solutions. Numerical calculations below instead
continue the low-dimensional pole equations directly, starting on
the physical branch.

Several limits give smaller algebraic problems. At $m=0$ the two
determinant poles merge and $G=\kappa/2$ on a differentiable branch.
The nontrivial response there is $\Gamma$, generically of degree
13, or degree five with one source. The finite-mass one-source
problem has degree 18, dropping to 12 in the neutral sector after
the degenerate cuts are removed. In the neutral symmetric case
there is just a cubic. These reductions and their exceptional
charts are detailed in Appendix~\ref{app:elimination}. Appendix
\ref{sec:fullsym} specifies the surviving lifts of the source-free
argument group; those algebraic transformations are not used to
select the physical saddle.

For clarity, a polynomial in the response is not obtained
by identifying it with the endpoint parameter. The two pole
equations solve linearly for $1/r$ and $1/s$ in terms of
$p,L$. After the endpoint squares are imposed the two expressions
for $L$ must have a common root. Their resultant gives a
polynomial $P(p)$; at generic parameters a linear subresultant
then determines $L$ from $p$. This produces a rational function
$g(p)$ for $G$. Equivalently, on the locus where the rational
reconstruction is regular, the response equation is the
monic product
\[
 \mathcal P_G(Y)=\prod_{P(p_i)=0}\bigl(Y-g(p_i)\bigr).
\]
The construction does not require choosing all 38 solutions
numerically: an algebraic norm computes the same product.
Removing the extraneous vanishing factors at the origin and
showing that the rational map does not reduce the generic
degree are separate steps, carried out in
Appendix~\ref{app:elimination}. In particular a degree bound on
the equation for $p$ would not, by itself, establish the degree
of $\mathcal P_G$.

\subsection{The degree-38 mass response}

For $h\ne0,\pm1$, solve the first two conditions for
$X=1/r$ and $Y=1/s$. The following polynomials make all denominators
explicit:
\begin{align}
 U_0&=-p^2(dp+B+ApL),& V_0&=p^2(e+pA)+BL,\\
 D&=\kappa(1-h^2)p^3,&
 X_0&=V_0-hU_0,& Y_0&=h(U_0-hV_0).
\end{align}
Thus $X=X_0/D$ and $Y=Y_0/D$. The endpoint constraints become two
cubics in $L$,
\begin{align}
 F_-(p,L)&=(p^2+2hL+h^2)X_0^2-D^2,\\
 F_+(p,L)&=(p^2+2L/h+h^{-2})Y_0^2-D^2 .
\end{align}
Up to a factor independent of $p$, the compact eliminant for the
full one-cut model is
\begin{equation}\label{eq:38}
 {\mathcal P_{38}(p)=p^{-8}\Res_L(F_-,F_+)=0.}
\end{equation}
Multiplying $F_+$ by $h^2$ gives polynomial rather than Laurent
coefficients in $h$, with no change in the equation on this chart.

\paragraph{Why the degree is 38.}
For generic parameters the leading coefficients of the two cubics
in $L$ have degree six in $p$. As $p\to\infty$, each cubic has one
root of order $p^2$ (respectively $-p^2/(2h)$ and $-hp^2/2$) and
two bounded roots. Each bounded pair tends to the zero of the
respective linear prefactor $X_0$ or $Y_0$ after division by $p^3$;
their two limits differ for $A\ne0$ and $h^2\ne1$. Five of the nine
cross-differences of roots have order $p^2$, and the other four tend
to nonzero constants. The product formula for the resultant gives
degree $3\cdot6+3\cdot6+5\cdot2=46$. At $p=0$, each cubic has
two roots of order $p^2$, with centers
$-(e+hB)p^2/B$ and $-(e+B/h)p^2/B$, respectively, and splitting
of order $p^3$. The two other roots tend to $-h/2$ and $-1/(2h)$.
Exactly four cross-differences vanish to order two. The leading
coefficients of the cubics do not vanish there generically, so the
resultant has a factor $p^8$ and the saturated degree is $46-8=38$.
The formula is for generic sources; singular slices should be
taken in the original pole conditions.

\paragraph{Equation for the mass generating function.}
On a generic solution of \eqref{eq:38}, the last nonconstant
subresultant in $L$ is linear, say $S_1(p)L+S_0(p)$. Recover
$L_*=-S_0/S_1$, then $r=D/X_0$ and $s=D/Y_0$ at $L=L_*$. The mass
derivative follows directly from differentiating the determinant:
\begin{equation}\label{eq:Gmass}
 G=\frac{\kappa}{2}
 \left[1-h\omega(-h)+h^{-1}\omega(-h^{-1})\right].
\end{equation}
For evaluation at either pole $z=-a$, let $b$ be the other positive
pole parameter and write $r_a=\mathcal R(-a)$, $r_b=\mathcal R(-b)$.
Subtracting the common simple pole in $V'$ and $y$ gives
\begin{align}
 2\omega(-a)={}&-A+\frac{B}{a^2}-\frac{e}{a}-\frac{\kappa}{b-a}
 -r_a\left(\frac A a+\frac{B}{p a^2}
 -\frac{\kappa}{a r_a}-\frac{\kappa b}{a r_b(b-a)}\right)
 -\frac{\kappa(a+L)}{r_a^2}.
\end{align}
Let $N_G(p)/D_G(p)$ denote \eqref{eq:Gmass} after the above
reconstruction. The monic polynomial in the mass GF variable $g$ is
\begin{equation}\label{eq:norm}
 {
 \frac{\Res_p\bigl(\mathcal P_{38},\,gD_G-N_G\bigr)}
      {\Res_p\bigl(\mathcal P_{38},D_G\bigr)}=0.
 }
\end{equation}
Equivalently, it is the determinant of $gI$ minus multiplication by
$N_GD_G^{-1}$ on the quotient algebra by $\mathcal P_{38}$.
The same construction applies to the two loop moments, extracted from
$\omega(z)=z^{-1}+M_+z^{-2}+\cdots$ at infinity and
$\omega(0)=-M_-$ on the corresponding continuation at zero.

For example, the expansion of \eqref{eq:y} at infinity gives
\begin{equation}
 2 M_+= B+\kappa(h+h^{-1})+\frac A2(p^2-L^2)
 +L\left(\frac Bp+\kappa\frac hr+\frac{\kappa}{hs}\right)
 +\kappa\frac{h^2}{r}+\frac{\kappa}{h^2s}.
\end{equation}
The other moment follows by exchanging $A,B$, sending $v\to-v$,
and replacing $(p,L,r,s)$ by
$(p^{-1},L/p^2,-hs/p,-r/(hp))$. The planar logarithm on a continued
branch can be reconstructed by integrating
$\Gamma(\lambda A,\lambda B;h)/\lambda$ along a source ray from
a reference point on that branch.

The degree is attained, including for $G$. For the exact
specialization $(\kappa,u,h,A,B)=(1,1/2,1/10,1/20,3/100)$,
the degree-38 factor is irreducible modulo $347$. In its residue
field the reconstructed $G$ satisfies $G^{347^{38}}=G$, but is
fixed by neither the $347^2$ nor the $347^{19}$ Frobenius power.
These are the two maximal proper subfields, so $G$ has degree 38
in this specialization. Together with \eqref{eq:38}, this
establishes the generic degree.

\subsection{Massless slice: \texorpdfstring{$m=0$ and $h=1$}{m=0 and h=1}}

The two poles coincide at $-1$, with $r=s$ on the regular
coalescing-pole sheet. Use $p=P/Q$ and set
\begin{align}
 C&=AP^3+BQ^3,& T&=(A+d)P+(B+e)Q,\\
 H_4&=B^2Q^4+dBPQ^3-eAP^3Q-A^2P^4,&
 J&=(P^2+Q^2)C-2P^2Q^2T .
\end{align}
Solving the two linear pole conditions for $L$ and $1/r$ and then
imposing $r^2=p^2+2L+1$ gives the binary form
\begin{equation}
 {\mathcal P_{13}(P,Q)=JH_4^2-4\kappa^2P^2Q^2C^3=0.}
\end{equation}
On $Q=1$ one has $L=-p^2T/C$ and $r=-2\kappa pC/H_4$. For the
regular interior sector, the source-free seed is $p=-e/d$ and
$r=-2\kappa/d$; use the pole conditions there, since the eliminant
degenerates at $A=B=0$. Charge conjugation acts by
$(A,B,v,[P:Q])\mapsto(B,A,-v,[Q:P])$. In the chart $A\ne B$,
$\chi=(p+1)/[(A-B)(p-1)]$ is invariant. The polynomial
\[
 \frac{\mathcal P_{13}(1+\delta\chi,-1+\delta\chi)}{\delta},
 \qquad \delta=A-B,
\]
is invariant under charge conjugation, so its coefficients can be
expressed in $\kappa,A+B,\delta^2,\delta v,v^2$. The binary form also
works on the symmetric source diagonal where that chart is singular.

The determinant inversion identity is
$Z_q(h)=h^{2NN_f} Z_q(1/h)$ at fixed $A,B$. Thus $G=\kappa/2$
at $h=1$ on a differentiable branch in the present normalization.
The nontrivial massless GF is $\Gamma=A M_++B M_-$,
whose polynomial is the norm construction \eqref{eq:norm}
with $\mathcal P_{13}(p,1)$ and with the rational expression
for $\Gamma$ in place of $G$. Its generic degree is 13.
For example at $(\kappa,u,A,B)=(1,1/2,1/5,1/3)$ the endpoint
polynomial is irreducible modulo 11 and $\Gamma$ is nonconstant
in the residue field.

\subsubsection{One source: the projective quintic}

Take $A=0$, $B\ne0$. Then $C=BQ^3$ and
$H_4=BQ^3(BQ+dP)$, so the factors supported on $Q=0$
can be removed before going to the affine chart.
Introduce the projective ratios and endpoint coordinate
\begin{equation}
 \alpha=\frac Bd,\qquad \beta=\frac ed,\qquad
 \gamma=\frac{2\kappa}{d},\qquad x=\frac{dp}{B}.
\end{equation}
These are coordinates on the chart $d\ne0$ of the homogeneous
coupling data $[B:d:e:2\kappa]$ after dehomogenization. For
$B=0$, $A\ne0$, use $(\alpha,\beta,\gamma,x)=(A/e,d/e,2\kappa/e,
e/(Ap))$ instead. Thus the same ratios and the same $x$ are used
on the two charts exchanged by charge conjugation.

With $x=X/W$ the exact binary quintic is
\begin{equation}\label{eq:one-massless}
 {(X+W)^2\bigl[W^3-\alpha(\alpha+2\beta)X^2W
                 -2\alpha^2X^3\bigr]-\gamma^2X^2W^3=0 .}
\end{equation}
In the $A=0$ chart the eliminated variables are reconstructed as
$L=-p^2(dp+e+B)/B$ and $r=-2\kappa p/(B+dp)$.
The normalized source derivative $\mathfrak g=\Gamma/d^2$
then has the explicit value
\begin{equation}\label{eq:one-source-response}
 {4\mathfrak g=x^{-2}-3\alpha^2x^2
 -4\alpha(\alpha+\beta)x-(\alpha^2+4\alpha\beta+\beta^2).}
\end{equation}
For the conjugate chart the normalization is $\mathfrak g=\Gamma/e^2$.
Equations \eqref{eq:one-massless} and \eqref{eq:one-source-response}
give a compact equation for the GF itself: take the resultant
of \eqref{eq:one-massless} at $W=1$ with
\[
 3\alpha^2x^4+4\alpha(\alpha+\beta)x^3
 +(4\mathfrak g+\alpha^2+4\alpha\beta+\beta^2)x^2-1
\]
in $x$, and normalize to be monic in $\mathfrak g$. Its generic
degree is five. The real branch is chosen by continuation of
the signed endpoint data, as in the two-source case.

\subsection{Neutral sector: \texorpdfstring{$u=0$}{u=0} and the symmetric-source cubic}

At $v=0$ charge conjugation still exchanges $A$ and $B$.
Thus the neutral slice alone does not imply reciprocal endpoint data
for unequal sources. In fact, even at $h=1$ the degree of $\Gamma$
remains 13 generically on this slice: $(\kappa,A,B)=(1,1/5,1/3)$
gives the prime-degree witness modulo 19.

On the real diagonal $A=B=g_0\ge0$, the neutral full-support branch
for $0<h<1$ has
\begin{equation}
 \Psi_0=g_0^2+2\kappa h g_0-\kappa^2\log(1-h^2),
 \qquad M_+=M_-=g_0+\kappa h.
\end{equation}
It reaches an opening gap at $g_0+\kappa h/(1-h)=1/2$.
On the real symmetric one-cut branch the reciprocal data are
$p=-1$, $s=r/h$. With $t=-r/(1+h)$ and
$\alpha=g_0(1+h)^2/(2h)$ the pole conditions reduce to
\begin{equation}
 {\alpha t^3+(1+\kappa-\alpha)t-\kappa=0.}
\end{equation}
The endpoint of the gapless branch has $L=-1$ and hence
$t=(1-h)/(1+h)$; putting this value in the cubic reproduces the
same gap condition. The $g_0=0$ limit recovers the gapped neutral
determinant-only branch. More generally, switching off both
sources first returns the source-free one-dimensional QCD problem;
its mass derivative is described by the source-free sextic on the
regular one-cut branch.

\subsubsection{One source at finite mass: the neutral degree-12 factor}

Set $A=0$, $B\ne0$ and $0<h<1$. Use $F_\pm$ from the full model.
The one-source eliminant
\[
 \mathcal R_{18}(p)=p^{-8}\Res_L(F_-,F_+)\big|_{A=0}
\]
has generic degree 18. Indeed, at infinity the two cubics in $L$
have constant leading coefficients and each has two roots growing
as $p^3$ and one as $p^2$. The two $p^3$ centers have distinct
leading terms $-hdp^3/B$ and $-dp^3/(hB)$; the two $p^2$ roots
have distinct leading coefficients. Thus eight cross-differences of
roots have order $p^3$ and one has order $p^2$, giving degree 26
before removing the same $p^8$ factor at the origin.

In the neutral sector set $d=e=d_0=1+\kappa$. There are
two degenerate components with $L=p$, where the two cut endpoints
coincide. The pole equations on these components give
$r=\epsilon(p+h)$ and $s=-\epsilon(p+h^{-1})$, with
$\epsilon=\pm1$. The remaining condition is
\begin{equation}
 C_\epsilon(p)=(B+d_0 p)(p+h)(1+hp)
                   +\epsilon\kappa(h^2-1)p^2=0 .
\end{equation}
For generic parameters these cubics are distinct. Discarding those
components gives the exact equation for the nondegenerate one-cut
neutral slice:
\begin{equation}\label{eq:neutral-one-source}
 {\mathcal P_{12}(p)=
 \frac{\mathcal R_{18}(p)|_{v=0}}{C_{+}(p)C_{-}(p)}=0.}
\end{equation}
Charge conjugation transports this to the $B=0$ chart using
$p\mapsto p^{-1}$. The monic equation for $G$ follows from
\eqref{eq:norm} with $\mathcal P_{12}$ in place of $\mathcal P_{38}$.
At $(\kappa,B,h)=(1,1/5,1/3)$ the residual factor is irreducible
modulo $7$ and the reconstructed $G$ has degree 12 in that field,
as checked by its Frobenius orbit. At the same parameters with
$v=1/2$, the one-source eliminant and $G$ have degree 18, with a
witness modulo $29$. These tests certify the generic degrees on
the respective slices. The degree-13 witness in the preceding
neutral discussion has two nonzero sources and $h=1$.

\section{Projective symmetry with independent sources}
\label{sec:fullsym}

As in Appendix~\ref{app:elimination}, $d$ denotes the pole parameter $d_{\mathrm p}$. The homogeneous flavor--charge triple is the same as in source-free
one-dimensional QCD:
\begin{equation}
 [a:b:c]=[N_f:N+N_f:q]=[\kappa:\kappa+1:v].
\end{equation}
For the model with sources we carry two additional coordinates,
$[a:b:c:A:B]$, with the common projective scale acting on both
sources. The physical chart is $b-a=1$; $h$ is a separate parameter.
In the source-free model \cite{VolProjective}, the signed permutations of $(a,b,c)$
modulo their common sign give $(\mathbb Z_2)^2\rtimes S_3\simeq S_4$.
With sources, the stabilizer of the coordinate position $a$ lifts
to the full algebraic curve. Its eight elements form $D_4$ (of order
eight). The following generators fix $h$:
\[
\begin{array}{c|c|c}
 & [a:b:c:A:B]\text{ before restoring }b-a=1
 & (p,L,r,s)\text{ after transformation}\\ \hline
 H & [-a:b:c:A:B] & (p,L,-r,-s)\\
 T & [a:c:b:-A:B] & (-p,L,-r,-s)\\
 C & [a:b:-c:B:A] & (p^{-1},L/p^2,-hs/p,-r/(hp))
\end{array}
\]
Here $H$ is the involution $\xi\mapsto\xi^{-1}$, with
$\xi=2\kappa+1$:
$(\kappa,v,A,B)\mapsto(-\kappa,v,A,B)/\xi$.
For $T$, writing $\Delta=v-\kappa$, the affine map is
$(\kappa,v,A,B)\mapsto(\kappa,\kappa+1,-A,B)/\Delta$.
In particular, $T$ changes the relative sign of the two sources but
does not exchange them. Substitution in the pole conditions and
the endpoint squares verifies all three lifts. Projectively
$T^2=C^2=H^2=1$ and $(TC)^2=H$, with $H$ central.

To express the symmetry directly on the mass generating function, set
$\zeta=(1-h^2)/(1+h^2)$ and $E=a^2+b^2+c^2$. In the physical chart,
\begin{equation}\label{eq:full-invariant}
 {
 W_{\rm full}=
 \frac{\zeta\bigl[G-\kappa/2-
   \kappa(h-h^{-1})(A+B)/4\bigr]-\kappa^2/2}{E}.}
\end{equation}
Under $H$ and $T$ the square bracket scales by the square of the
projective chart factor, just as $E$ and $\kappa^2$ do. This follows by
taking the finite parts at the determinant poles in \eqref{eq:y}:
the changed signs of $r,s$ must be used when evaluating $G$.
Charge conjugation $C$ leaves the bracket and $E$ fixed.
At $A=B=0$, \eqref{eq:full-invariant} is exactly the source-free
normalization. Applying the norm construction \eqref{eq:norm}
to $W_{\rm full}$ gives a monic degree-38 polynomial invariant under
the lifted $D_4$. Charge conjugation is an integral identity;
$H$ and $T$ are algebraic transformations of the transported
branches and need not stay in the physical region of $\kappa$ or
on the physical sheet. The independent mass inversion $h\mapsto h^{-1}$
also preserves $W_{\rm full}$: the square bracket and $\zeta$ both change
sign, by the determinant inversion identity. This inversion is not
counted among the eight lifts of the source-free argument group.

The massless limit distinguishes this $D_4$ from the larger $S_4$.
For generic finite nonzero sources, the monic polynomial for
$W_{\rm full}$ tends as $h\to1$ to
\begin{equation}\label{eq:full-sym-massless}
 w^{23}\left(w+\frac{a^2}{2E}\right)^{13}
        \left(w+\frac{b^2}{2E}\right)
        \left(w+\frac{c^2}{2E}\right).
\end{equation}
The 13 regular branches with $r=s\ne0$ are the branches of the
massless endpoint equation and give $w=-a^2/(2E)$. The other two
nonzero limits come from singular branches with $p\to+1$ and
$p\to-1$. To see this, put $\delta=h^{-1}-h$ and
$F=AL+B/p+d+h(A+BL/p^3+e/p)$. The combined pole conditions give
$F+\kappa\delta/s=0$. On the two branches
$\delta/s\to-F_0/\kappa$, where $F_0=d+e=2b$ or $d-e=-2c$,
respectively. The scaled system in $r/\delta,s/\delta,p,L$
has one simple solution at each sign, with $L\to-1$ and
$r/\delta\to F_0/\kappa$. The finite-part formula for
\eqref{eq:Gmass} then gives $w=-F_0^2/(8E)$.

For completeness, no endpoint parameter $p$ escapes to infinity:
the pole equations would force $L$ of order $p^2$ and both $r,s$
of order $\delta/p^2$, in conflict with \eqref{eq:quadric}.
Charge conjugation rules out $p\to0$, and then the first pole equation
rules out $L\to\infty$. The remaining branches either have
$r\to-s\ne0$ or have $r,s\to0$ with $p\not\to\pm1$.
In the latter case the bounded sum $1/r+1/s$ and \eqref{eq:quadric}
give $\delta=O(r^3)$, so terms proportional to $\delta/r^2$ vanish.
In both cases $r/s\to-1$ and
$\delta(G-\kappa/2)\to\kappa^2$. Since $\zeta/\delta\to1/2$,
these branches give $w=0$, accounting for the remaining 23 roots.

The multiplicities $13,1,1$ distinguish the position $a$.
Any of the 16 signed permutations outside $D_4$ moves that position
and therefore moves the unique root of multiplicity 13. Thus $D_4$
is the maximal subgroup of the source-free $S_4$ admitting a
regular lift that preserves \eqref{eq:full-invariant}, the massless
boundary, and generic finite sources there. This argument does not
exclude singular correspondences or transformations between
different boundary strata. At $h=1$ the invariant itself no longer
resolves the 13 regular branches; $\Gamma$ remains the useful
generating function on that slice.

\section{Derivation of the local source Hessian}\label{app:hessian}
This appendix uses the endpoints and variables of
Section~\ref{sec:screening} and Eq.~\eqref{eq:screen-geometry}.
To derive the Hessian, linearize $\omega_++\omega_-=V'$ on the cut
at fixed $h,\kappa$. The variations have at most inverse square
root singularities at the endpoints, are regular at zero, and
decay as $O(z^{-2})$ at infinity. These conditions give
\begin{align}
 \partial_A\omega&=-\tfrac12+\frac{z-L}{2\mathcal R},&
 \partial_B\omega&=\frac1{2z^2}-
      \frac{p/z^2-L/(pz)}{2\mathcal R},&
 \partial_v\omega&=\frac1{2z}-\frac{z+p}{2z\mathcal R},
 \label{eq:screen-linear-loop}
\end{align}
where $\mathcal R$ is the square root from Section~\ref{sec:fixed-sector}.
Expansion at infinity and zero gives
\begin{equation}\label{eq:screen-fixed-hess}
 \Psi_{(A,B)(A,B)}=
 \begin{pmatrix}-p^2\sigma(1-\sigma)&\sigma\\
 \sigma&-p^{-2}\sigma(1-\sigma)\end{pmatrix},\qquad
 \Psi_{(A,B)v}=\begin{pmatrix}-p\sigma\\\sigma/p\end{pmatrix}.
\end{equation}
The logarithmic moment of the last variation in
\eqref{eq:screen-linear-loop}, evaluated on the arc
$z=(-p)e^{i\theta}$, gives
\begin{equation}
 -\Psi_{vv}=\frac1\pi\int_0^{\theta_0}
 \frac{\theta\sin(\theta/2)}{\sqrt{\sigma-\sin^2(\theta/2)}}\,\dd\theta
 =\mathcal L .
 \label{eq:screen-sector-curvature}
\end{equation}
To evaluate the integral, set $c_0=\sqrt{1-\sigma}$ and integrate
by parts. The result is
\[\frac4\pi\int_0^{\arccos c_0}\operatorname{arcosh}(\cos t/c_0)\,\dd t.\]
Its derivative with respect to $c_0$ is $-2/c_0$ and it vanishes
at $c_0=1$.

\section{Spatial simulation and screening estimators}\label{app:MC}

\paragraph{Action and observables.} The simulation uses the weight of Eq.~\eqref{eq:spatial-action}:
\[
 Z=\int\prod_i dU_i\,
 \exp\!\left[\frac b3\sum_{\langle ij\rangle}
       \Ree(\Tr U_i\,\Tr U_j^\dagger)\right]
 \prod_i |\det(1+e^{-m+i\theta}U_i)|^{2N_f}.
\]
Here $N=N_f=3$, $m=0$, $b=0.65$ and
$\theta=\phi/3$. This is the massless value of the static
parameter, with no spatial quark propagation. At each site
two eigenangles were sampled, with the third fixed by
$\det U=1$ and the squared Haar Vandermonde included in the
Metropolis weight. We used periodic sizes $L=6,10$, four
chains for each angle and size, and 85\,000 ($L=6$) or
65\,000 ($L=10$) production sweeps per chain, following
4\,000 warmup sweeps. At $\phi=\pi$ the two selected
orientations were sampled separately and reflected into the
same phase before computing connected observables.

\paragraph{Correlator estimator.} The real two-component spin is
$O=(\Ree\mathrm w_\theta,\Imm\mathrm w_\theta)$, where
$\mathrm w_\theta=e^{i\theta}\Tr U/3$. We use connected wall correlators
\[
 C_{ab}(r)=L^2\,\langle \overline O_a(z)\,\overline O_b(z+r)\rangle_c,
 \qquad
 \overline O_a(z)=L^{-2}\sum_{\mathbf x_\perp} O_a(\mathbf x_\perp,z),
\]
averaged over directions and origins. For $r=1$ one spin is
replaced by its conditional one-site mean (symmetrized over
the endpoints); for $r\ge2$ both are replaced. The latter
replacement is an exact conditional identity because the two
sites have no direct bond at these separations. The conditional
means are tabulated by independent two-angle quadrature and
interpolated; errors were below $4\times10^{-8}$ at test points
in the populated source windows.

We determine a common basis from
$C(1)v=\lambda C(0)v$ and define the effective masses through
the periodic cosh ratio, using either $C(1)/C(0)$ or the projected
$C(2)/C(1)$. Using a fresh generalized eigenbasis from $C(1),C(2)$
changes the central masses by less than $0.003$.
Statistical errors were estimated by resampling blocks of 4,000 sweeps. A separate pilot tested the pressure slope with three chains at interior angles and independent initializations in each phase at the RW boundary. The source archive includes the scripts and blocked measurements for both calculations.

\paragraph{Reproducibility.}
The accompanying source archive contains the one-site continuation,
the degree certificate, the blocked spatial measurements and
the scripts used to analyze and plot them.

\section*{Acknowledgements}
The author thanks O.~Borisenko and V.~Chelnokov for stimulating
discussions; this work was supported by the Simons Foundation
(Grant \mbox{SFI-PD-Ukraine-00014578}).


\begin{thebibliography}{99}
\bibitem{VolSU}
S.~Voloshyn, \emph{Polyakov loop model with exact static quark determinant in the 't Hooft--Veneziano limit: SU(N) case},
\href{https://arxiv.org/abs/2512.07026}{arXiv:2512.07026 [hep-lat]} (earlier version).

\bibitem{VolU}
S.~Voloshyn, \emph{Polyakov loop model with exact static quark determinant in the 't Hooft--Veneziano limit: U(N) case},
Phys. Rev. D \textbf{113}, 074526 (2026),
\href{https://doi.org/10.1103/5mdx-j8qy}{doi:10.1103/5mdx-j8qy}.

\bibitem{QCDone}
O.~Borisenko, V.~Chelnokov, S.~Voloshyn and P.~Yefanov,
\emph{One-dimensional QCD at finite density and its 't Hooft--Veneziano limit},
JHEP \textbf{01}, 008 (2025), \href{https://arxiv.org/abs/2410.02328}{arXiv:2410.02328}.

\bibitem{VolProjective}
S.~Voloshyn, \emph{Projective symmetries and differential recursions for
the large-$N$ Gross--Witten--Wadia and one-dimensional QCD matrix models},
\href{https://arxiv.org/abs/2611.11670v1}{arXiv:2611.11670v1 [math-ph]} (2026).

\bibitem{BCVlarge}
O.~Borisenko, V.~Chelnokov and S.~Voloshyn, \emph{The large $N$ limit of SU(N) integrals in lattice models},
Nucl. Phys. B \textbf{960}, 115177 (2020), \href{https://arxiv.org/abs/2008.00773}{arXiv:2008.00773}.

\bibitem{BCVphase}
O.~Borisenko, V.~Chelnokov and S.~Voloshyn, \emph{Polyakov loop models in the large $N$ limit: Phase diagram at finite density},
Phys. Rev. D \textbf{105}, 014501 (2022), \href{https://arxiv.org/abs/2111.00474}{arXiv:2111.00474}.

\bibitem{BCVscreening}
O.~Borisenko, V.~Chelnokov and S.~Voloshyn, \emph{Polyakov loop models in the large $N$ limit: Correlation function and screening masses},
Phys. Rev. D \textbf{109}, 094503 (2024),
\href{https://doi.org/10.1103/PhysRevD.109.094503}{doi:10.1103/PhysRevD.109.094503}.

\bibitem{Russo}
J.~G.~Russo, \emph{Phases of unitary matrix models and lattice QCD2},
\href{https://arxiv.org/abs/2010.02950}{arXiv:2010.02950}.

\bibitem{RussoTierz}
J.~G.~Russo and M.~Tierz, \emph{Multiple phases in a generalized Gross--Witten--Wadia model},
\href{https://arxiv.org/abs/2007.08515}{arXiv:2007.08515}.

\bibitem{SantilliTierz}
L.~Santilli and M.~Tierz, \emph{Exact equivalences and phase discrepancies between random matrix ensembles},
\href{https://arxiv.org/abs/2003.10475}{arXiv:2003.10475}.

\bibitem{Malik}
A.~Malik, \emph{Large-$N$ dynamics of a QCD-inspired unitary matrix model},
\href{https://arxiv.org/abs/2604.18182}{arXiv:2604.18182 [hep-th]} (2026).

\bibitem{MalikAhmed}
A.~Malik and A.~Ahmed, \emph{Spectral and phase structure of a QCD-inspired unitary matrix model with Fisher--Hartwig singularities},
\href{https://arxiv.org/abs/2601.02953}{arXiv:2601.02953 [hep-th]} (2026).

\bibitem{BCVdual}
O.~Borisenko, V.~Chelnokov and S.~Voloshyn, \emph{Dual formulations of Polyakov loop lattice models},
Phys. Rev. D \textbf{102}, 014502 (2020), \href{https://arxiv.org/abs/2005.11073}{arXiv:2005.11073}.

\bibitem{Fromm}
M.~Fromm, J.~Langelage, S.~Lottini and O.~Philipsen, \emph{The QCD deconfinement transition for heavy quarks and all baryon chemical potentials},
JHEP \textbf{01}, 042 (2012), \href{https://arxiv.org/abs/1111.4953}{arXiv:1111.4953}.

\bibitem{DualPhase}
O.~Borisenko, V.~Chelnokov, E.~Mendicelli and A.~Papa,
\emph{Dual simulation of a Polyakov loop model at finite baryon density: Phase diagram and local observables},
Nucl. Phys. B \textbf{965}, 115332 (2021), \href{https://arxiv.org/abs/2011.08285}{arXiv:2011.08285}.

\bibitem{GrossWitten}
D.~J.~Gross and E.~Witten, \emph{Possible third-order phase transition in the large-$N$ lattice gauge theory},
Phys. Rev. D \textbf{21}, 446 (1980).

\bibitem{Wadia}
S.~R.~Wadia, \emph{$N=\infty$ phase transition in a class of exactly soluble model lattice gauge theories},
Phys. Lett. B \textbf{93}, 403 (1980).

\bibitem{RW}
A.~Roberge and N.~Weiss, \emph{Gauge theories with imaginary chemical potential and the phases of QCD},
Nucl. Phys. B \textbf{275}, 734 (1986).

\bibitem{GWWnote}
\emph{Exact SU(N) GWW input and the susceptibility method for Polyakov-loop phases, correlations and screening masses},
unpublished working manuscript, 24 September 2026, Secs.~8 and 10.

\bibitem{DualMass}
O.~Borisenko, V.~Chelnokov, E.~Mendicelli and A.~Papa,
\emph{Dual simulation of a Polyakov loop model at finite baryon density: Correlations and screening masses},
Nucl. Phys. B \textbf{998}, 116424 (2024), \href{https://arxiv.org/abs/2309.06104}{arXiv:2309.06104}.

\bibitem{Kashiwa}
K.~Kashiwa and R.~D.~Pisarski, \emph{The Roberge--Weiss transition and 't Hooft loops},
\href{https://arxiv.org/abs/1301.5344}{arXiv:1301.5344}.

\bibitem{deFP}
P.~de Forcrand and O.~Philipsen, \emph{Constraining the QCD phase diagram by tricritical lines at imaginary chemical potential},
Phys. Rev. Lett. \textbf{105}, 152001 (2010), \href{https://arxiv.org/abs/1004.3144}{arXiv:1004.3144}.

\bibitem{Andreoli}
M.~Andreoli, C.~Bonati, M.~D'Elia, M.~Mesiti, F.~Negro, A.~Rucci and F.~Sanfilippo,
\emph{Gauge-invariant screening masses and static quark free energies in $N_f=2+1$ QCD at nonzero baryon density},
Phys. Rev. D \textbf{97}, 054515 (2018), \href{https://arxiv.org/abs/1712.09996}{arXiv:1712.09996}.
\end{thebibliography}
\end{document}